# A simple and accurate algorithm for path integral molecular dynamics with the Langevin thermostat

Jian Liu [1, a)], Dezhang Li [1, b)], Xinzijian Liu [1, b)]

1.   *Beijing National Laboratory for Molecular Sciences, Institute of Theoretical and Computational Chemistry, College of Chemistry and Molecular Engineering, Peking University, Beijing 100871, China*

a)   Electronic mail: jianliupku@pku.edu.cn

b)   Both authors contributed equally to the work

We introduce a novel simple algorithm for thermostatting path integral molecular dynamics (PIMD) with the Langevin equation. The staging transformation of path integral beads is employed for demonstration. The optimum friction coefficients for the staging modes in the free particle limit are used for all systems. In comparison to the path integral Langevin equation (PILE) thermostat, the new algorithm exploits a different order of splitting for the phase space propagator associated to the Langevin equation. While the error analysis is made for both algorithms, they are also employed in the PIMD simulations of three realistic systems (the $H_2O$ molecule, liquid *para*-hydrogen, and liquid water) for comparison. It is shown that the new thermostat increases the time interval of PIMD by a factor of 4~6 or more for achieving the same accuracy. In addition, supplemental material shows the error analysis made for the algorithms when the normal-mode transformation of path integral beads is used.

## I. Introduction

In 1953 Feynman first presented imaginary time path integral to study liquid Helium[1], which already demonstrated the mapping of a quantum system onto a classical model consisting of the Feynman ring of "beads" (i.e., replicas of the system) connected by harmonic springs[1]. In 1981 Chandler and Wolynes then suggested the quantum-classical isomorphism that established the relationship between quantum concepts and the classical polymer language[2, 3]. Imaginary time path integral has not only provided a physical picture but also offered a powerful computational framework for studying quantum statistical effects[3-5]. Such effects (including zero point energy, tunneling, and quantum exchange effects) become important at low temperatures and/or in realistic systems that contain light atoms such as hydrogen or helium.

In 1984 Parrinello and Rahman proposed that artificial momenta could be assigned to the mapping polymer such that molecular dynamics (MD) could be employed to perform the path integral sampling[6]. The normal-mode transformation[7-9] or staging transformation[10, 11] was introduced to deal with the stiffness of the harmonic springs between the beads. Thermostatting methods such as the Andersen thermostat[12], Langevin dynamics[13, 14], Nosé-Hoover chain (NHC)[10, 11], *etc.* were implemented in PIMD to ensure a proper canonical distribution of the path integral beads. As it is often not a trivial task to adjust moves of path integral Monte Carlo (PIMC) for general molecular systems, path integral molecular dynamics (PIMD) thus offers a more convenient computational technique for simulating structural and thermodynamic properties when quantum exchange effects are not significant.

As early as in 1980s Langevin dynamics was already introduced for thermostatting PIMD by Gillan[13] and by Singer and Smith[14]. It has also been investigated by Müser *et al.*[15] and by Drozdov and Talkner[16]. More recently, Ceriotti *et al.* have developed a path integral Langevin equation (PILE) thermostat that combines a simple (white noise) Langevin thermostat with the velocity Verlet algorithm to give an efficient sampling of the canonical distribution for PIMD[17]. It is demonstrated that in terms of sampling efficiency PILE is comparable to the NHC thermostat[10] for PIMD[17]. The implementation of PILE is very straightforward for general molecular systems. These suggest that it is worth investigating more stochastic methods for PIMD.

The purpose of this paper is to present a novel, simple, and accurate algorithm for accomplishing PIMD with Langevin thermostats. Section II first briefly reviews PIMD with the staging transformation of path integral beads. After demonstrating how the PILE thermostat[17] can be implemented for staging PIMD, we introduce a more accurate and robust integrator for propagating PIMD with the (white noise) Langevin thermostat. Section III applies both integrators to three typical realistic molecular systems, namely, the water molecule, liquid *para*-hydrogen, and liquid water. The two integrators are compared by studying two thermodynamic properties, the average kinetic energy (obtained by either the primitive or virial estimator) and the average potential energy. The performance is then investigated as a function of the time interval of PIMD. (More discussions are given in Appendices I-III and in Supplemental Material[18].) Conclusions and outlook follow in Section IV.

## II. Theory

### 1. Thermodynamic properties

Any thermodynamic property of the canonical ensemble is of the general form

$$\left\langle \hat{B} \right\rangle = \frac{1}{Z} \text{Tr}\left( e^{-\beta \hat{H}} \hat{B} \right) \ , \tag{1}$$

where $Z = \text{Tr}\left[ e^{-\beta \hat{H}} \right]$ $\left( \beta = 1/k_B T \right)$ is the partition function, $\hat{H}$ the (time-independent) Hamiltonian of the system with the total number of degrees of freedom $N$, which we assume to be of standard Cartesian form

$$\hat{H} = \frac{1}{2} \hat{\mathbf{p}}^T \mathbf{M}^{-1} \hat{\mathbf{p}} + V\left( \hat{\mathbf{x}} \right) \ , \tag{2}$$

where $\mathbf{M}$ is the diagonal 'mass matrix' with elements $\{m_j\}$, and $\hat{\mathbf{p}}$ and $\hat{\mathbf{x}}$ are the momentum and coordinate operators, respectively; and $\hat{B}$ is an operator relevant to the specific property of interest.

Express Eq. (1) in the coordinate space $\mathbf{x}$, i.e.,

$$\left\langle \hat{B} \right\rangle = \frac{\int d\mathbf{x} \left\langle \mathbf{x} \left| e^{-\beta \hat{H}} \hat{B} \right| \mathbf{x} \right\rangle}{\int d\mathbf{x} \left\langle \mathbf{x} \left| e^{-\beta \hat{H}} \right| \mathbf{x} \right\rangle} \ . \tag{3}$$

Inserting path integral beads to evaluate the term $\left\langle \mathbf{x} \left| e^{-\beta \hat{H}} \right| \mathbf{x} \right\rangle$ leads to

$$\begin{aligned} Z &= \int d\mathbf{x} \left\langle \mathbf{x} \left| e^{-\beta \hat{H}} \right| \mathbf{x} \right\rangle \\ &\overset{\mathbf{x}_1 = \mathbf{x}}{=} \lim_{P \to \infty} \int d\mathbf{x}_1 \int d\mathbf{x}_2 \cdots \int d\mathbf{x}_P \left( \frac{P}{2\pi\beta\hbar^2} \right)^{NP/2} |\mathbf{M}|^{P/2} \\ &\quad \times \exp\left\{ -\frac{P}{2\beta\hbar^2} \sum_{i=1}^{P} \left[ \left( \mathbf{x}_{i+1} - \mathbf{x}_i \right)^T \mathbf{M} \left( \mathbf{x}_{i+1} - \mathbf{x}_i \right) \right] - \frac{\beta}{P} \sum_{i=1}^{P} V\left( \mathbf{x}_i \right) \right\} \end{aligned} \tag{4}$$

where $\mathbf{x}_{P+1} \equiv \mathbf{x}_1$ and $P$ is the number of path integral beads. Then the numerator of Eq. (3) becomes

$$\int d\mathbf{x} \langle \mathbf{x} | e^{-\beta \hat{H}} \hat{B} | \mathbf{x} \rangle \stackrel{\mathbf{x}_1 = \mathbf{x}}{=} \lim_{P \to \infty} \int d\mathbf{x}_1 \int d\mathbf{x}_2 \cdots \int d\mathbf{x}_P \left( \frac{P}{2\pi\beta\hbar^2} \right)^{NP/2} |\mathbf{M}|^{P/2}$$

$$\times \exp\left\{ -\frac{P}{2\beta\hbar^2} \sum_{i=1}^{P} \left[ (\mathbf{x}_{i+1} - \mathbf{x}_i)^T \mathbf{M} (\mathbf{x}_{i+1} - \mathbf{x}_i) \right] - \frac{\beta}{P} \sum_{i=1}^{P} V(\mathbf{x}_i) \right\} \quad . \qquad (5)$$

$$\times \tilde{B}(\mathbf{x}_1, \cdots, \mathbf{x}_P)$$

The denominator of Eq. (3) takes the same form as Eq. (5) for $\hat{B} = 1$. It is straightforward to show that the estimator $\tilde{B}(\mathbf{x}_1, \cdots, \mathbf{x}_P)$ for any coordinate dependent operator $\hat{B}(\hat{\mathbf{x}})$ is

$$\tilde{B}(\mathbf{x}_1, \cdots, \mathbf{x}_P) = \frac{1}{P} \sum_{j=1}^{P} B(\mathbf{x}_j) \quad . \qquad (6)$$

When $\hat{B} = \frac{1}{2} \hat{\mathbf{p}}^T \mathbf{M}^{-1} \hat{\mathbf{p}}$ is the kinetic energy operator, the primitive estimator is

$$\tilde{B}(\mathbf{x}_1, \cdots, \mathbf{x}_P) = \frac{NP}{2\beta} - \sum_{j=1}^{P} \frac{P}{2\beta^2\hbar^2} \left[ (\mathbf{x}_{j+1} - \mathbf{x}_j)^T \mathbf{M} (\mathbf{x}_{j+1} - \mathbf{x}_j) \right] \qquad (7)$$

and the virial version is

$$\tilde{B}(\mathbf{x}_1, \cdots, \mathbf{x}_P) = \frac{N}{2\beta} + \frac{1}{2P} \sum_{j=1}^{P} \left[ (\mathbf{x}_j - \mathbf{x}^*)^T \frac{\partial V(\mathbf{x}_j)}{\partial \mathbf{x}_j} \right] \quad , \qquad (8)$$

where

$$\mathbf{x}^* = \mathbf{x}_c \equiv \frac{1}{P} \sum_{j=1}^{P} \mathbf{x}_j \qquad (9)$$

or $\mathbf{x}^*$ can be any one of the $P$ beads

$$\mathbf{x}^* = \mathbf{x}_i \quad , \tag{10}$$

with $i$ fixed in Eq. (8). It was suggested that the virial estimator is numerically more favorable than the primitive one as the number of beads $P$ increases [8].

## 2. Path Integral Molecular Dynamics

Consider the staging transformation of Tuckerman *et al.* [10, 11, 19]

$$\begin{aligned} \xi_1 &= \mathbf{x}_1 \\ \xi_j &= \mathbf{x}_j - \frac{(j-1)\mathbf{x}_{j+1} + \mathbf{x}_1}{j} \qquad \left(j = \overline{2, P}\right) \end{aligned} \tag{11}$$

Its inverse transformation takes the following convenient recursive form

$$\begin{aligned} \mathbf{x}_1 &= \xi_1 \\ \mathbf{x}_j &= \xi_j + \frac{j-1}{j}\mathbf{x}_{j+1} + \frac{1}{j}\xi_1 \qquad \left(j = \overline{2, P}\right) \end{aligned} \tag{12}$$

Its close form can be expressed as

$$\begin{aligned} \mathbf{x}_1 &= \xi_1 \\ \mathbf{x}_j &= \xi_1 + \sum_{k=j}^{P} \frac{j-1}{k-1}\xi_l \qquad \left(j = \overline{2, P}\right) \end{aligned} \tag{13}$$

If one defines

$$\omega_P = \frac{\sqrt{P}}{\beta \hbar} \quad , \tag{14}$$

Eq. (4) becomes

$$Z = \lim_{P \to \infty} \left(\frac{P}{2\pi\beta\hbar^2}\right)^{NP/2} |\mathbf{M}|^{P/2} \int d\boldsymbol{\xi}_1 \int d\boldsymbol{\xi}_2 \cdots \int d\boldsymbol{\xi}_P$$
$$\times \exp\left\{-\beta \sum_{j=1}^{P} \left[\frac{1}{2}\omega_P^2 \boldsymbol{\xi}_j^T \overline{\mathbf{M}}_j \boldsymbol{\xi}_j + \frac{1}{P}V\left(\mathbf{x}_j\left(\boldsymbol{\xi}_1,\cdots,\boldsymbol{\xi}_P\right)\right)\right]\right\},$$

(15)

with the (diagonal) mass matrices given by

$$\overline{\mathbf{M}}_1 = 0$$
$$\overline{\mathbf{M}}_j = \frac{j}{j-1}\mathbf{M} \qquad \left(j = \overline{2,P}\right)$$

(16)

If one defines

$$\phi\left(\boldsymbol{\xi}_1,\cdots,\boldsymbol{\xi}_P\right) = \frac{1}{P}\sum_{j=1}^{P}V\left(\mathbf{x}_j\left(\boldsymbol{\xi}_1,\cdots,\boldsymbol{\xi}_P\right)\right) ,$$

(17)

one then obtains the chain rule

$$\frac{\partial\phi}{\partial\boldsymbol{\xi}_1} = \sum_{i=1}^{P}\frac{\partial\phi}{\partial\mathbf{x}_i} = \frac{1}{P}\sum_{i=1}^{P}V'\left(\mathbf{x}_i\right)$$
$$\frac{\partial\phi}{\partial\boldsymbol{\xi}_j} = \frac{\partial\phi}{\partial\mathbf{x}_j} + \frac{j-2}{j-1}\frac{\partial\phi}{\partial\boldsymbol{\xi}_{j-1}} \qquad \left(j = \overline{2,P}\right)$$

(18)

from Eqs. (11)-(12). Employing the isomorphism strategy proposed by Chandler and Wolynes[2], one can insert fictitious momenta $\left(\mathbf{p}_1,\cdots,\mathbf{p}_P\right)$ into Eq. (15), which leads to

$$Z = \lim_{P \to \infty} \left(\frac{P}{4\pi^2\hbar^2}\right)^{NP/2} |\mathbf{M}|^{P/2} \left(\prod_{j=1}^{P}\left|\tilde{\mathbf{M}}_j\right|\right)^{-1/2} \int \left(\prod_{j=1}^{P}d\boldsymbol{\xi}_j d\mathbf{p}_j\right)$$
$$\times \exp\left[-\beta H_{eff}\left(\boldsymbol{\xi}_1,\cdots,\boldsymbol{\xi}_P;\mathbf{p}_1,\cdots,\mathbf{p}_P\right)\right]$$

(19)

with the effective Hamiltonian given by

$$H_{eff}\left(\xi_1,\cdots,\xi_P;\mathbf{p}_1,\cdots,\mathbf{p}_P\right)=\sum_{j=1}^{P}\frac{1}{2}\mathbf{p}_j^T\tilde{\mathbf{M}}_j^{-1}\mathbf{p}_j+U_{eff}\left(\xi_1,\cdots,\xi_P\right)\quad,\qquad(20)$$

where

$$U_{eff}\left(\xi_1,\cdots,\xi_P\right)=\sum_{j=1}^{P}\frac{1}{2}\omega_P^2\xi_j^T\bar{\mathbf{M}}_j\xi_j+\phi\left(\xi_1,\cdots,\xi_P\right)\quad.\qquad(21)$$

The fictitious masses are chosen as

$$\begin{aligned}\tilde{\mathbf{M}}_1&=\mathbf{M}\\ \tilde{\mathbf{M}}_j&=\bar{\mathbf{M}}_j\qquad\left(j=\overline{2,P}\right)\end{aligned}\qquad(22)$$

such that all staging modes $\left(\xi_2,\cdots,\xi_P\right)$ will move on the same time scale. The thermodynamic property Eq. (3) is then expressed as

$$\left\langle\hat{B}\right\rangle=\lim_{P\to\infty}\frac{\int\left(\prod_{j=1}^{P}d\xi_j d\mathbf{p}_j\right)\exp\left\{-\beta H_{eff}\left(\xi_1,\cdots,\xi_P;\mathbf{p}_1,\cdots,\mathbf{p}_P\right)\right\}\tilde{B}\left(\mathbf{x}_1,\cdots,\mathbf{x}_P\right)}{\int\left(\prod_{j=1}^{P}d\xi_j d\mathbf{p}_j\right)\exp\left\{-\beta H_{eff}\left(\xi_1,\cdots,\xi_P;\mathbf{p}_1,\cdots,\mathbf{p}_P\right)\right\}}\quad.\qquad(23)$$

One can sample $\left(\xi_1,\cdots,\xi_P,\mathbf{p}_1,\cdots,\mathbf{p}_P\right)$ in a molecular dynamics (MD) scheme for evaluating the thermodynamic property. That is, Eq. (23) leads to

$$\begin{aligned}\dot{\xi}_j&=\tilde{\mathbf{M}}_j^{-1}\,\mathbf{p}_j\\ \dot{\mathbf{p}}_j&=-\omega_P^2\bar{\mathbf{M}}_j\xi_j-\frac{\partial\phi}{\partial\xi_j}\qquad\left(j=\overline{1,P}\right)\end{aligned}\quad.\qquad(24)$$

The equations of motion for $\left(\boldsymbol{\xi}_1, \cdots, \boldsymbol{\xi}_P, \mathbf{p}_1, \cdots, \mathbf{p}_P\right)$ in Eq. (24) must be coupled to a thermostatting method to ensure a proper canonical distribution for $\left(\boldsymbol{\xi}_1, \cdots, \boldsymbol{\xi}_P, \mathbf{p}_1, \cdots, \mathbf{p}_P\right)$. Note that only the configurational distribution of PIMD is important in Eq. (23).

### 3. Algorithms for PIMD with Langevin thermostats

When a simple (white noise) Langevin dynamics is employed to thermostat the staging path integral variables $\left(\boldsymbol{\xi}_1, \cdots, \boldsymbol{\xi}_P, \mathbf{p}_1, \cdots, \mathbf{p}_P\right)$ in PIMD, Eq. (24) becomes

$$\begin{pmatrix} \dot{\boldsymbol{\xi}}_j \\ \dot{\mathbf{p}}_j \end{pmatrix} = \begin{pmatrix} \tilde{\mathbf{M}}_j^{-1} \mathbf{p}_j \\ -\omega_P^2 \bar{\mathbf{M}}_j \boldsymbol{\xi}_j - \dfrac{\partial \phi}{\partial \boldsymbol{\xi}_j} - \gamma_{Lang}^{(j)} \mathbf{p}_j + \sqrt{\dfrac{2\gamma_{Lang}^{(j)}}{\beta}} \left(\tilde{\mathbf{M}}_j\right)^{1/2} \boldsymbol{\eta}_j(t) \end{pmatrix} \quad \left(j = \overline{1,P}\right), \qquad (25)$$

Here $\boldsymbol{\eta}_j(t)$ is a vector. Its (white-noise) element $\eta_j^i(t)$ is an independent Gaussian-distributed random number with zero mean and unit variance [ $\left\langle \eta_j^i(t) \right\rangle = 0$ and $\left\langle \eta_j^i(t)\eta_j^i(t') \right\rangle = \delta(t-t')$ ], which is different for each physical degree of freedom $\left(i = \overline{1,N}\right)$, each staging mode $\left(j = \overline{1,P}\right)$, and each time step. The Langevin friction coefficient $\gamma_{Lang}^{(j)}$ is the same for the staging modes $\left(j = \overline{2,P}\right)$ and all degrees of freedom $\left(i = \overline{1,N}\right)$ because they share the same frequency $\omega_P$. The optimum value for the friction coefficient $\gamma_{Lang}^{(j)}$ is

$$\gamma_{Lang}^{opt} = \omega_P \quad, \qquad (26)$$

which offers the most efficient configurational sampling in the free particle limit for the staging variables $\left(\boldsymbol{\xi}_2, \cdots, \boldsymbol{\xi}_P\right)$ [20, 21]. (See Appendix I.) Because the number of path integral beads $P$ in principle approaches infinity in PIMD for obtaining exact quantum thermodynamic properties [Eq. (23)], the optimum friction coefficient $\gamma_{Lang}^{opt}$ for overall sampling of the configurational distribution is then often considerably large for converged results according to Eqs. (14) and (26).

Because the harmonic force term $-\omega_P^2 \bar{\mathbf{M}}_j \boldsymbol{\xi}_j$ often varies much more frequently than the force term $-\dfrac{\partial \phi}{\partial \boldsymbol{\xi}_j}$, Eq. (25) can be divided into three parts

$$
\begin{pmatrix} \dot{\boldsymbol{\xi}}_j \\ \dot{\mathbf{p}}_j \end{pmatrix} = \underbrace{\begin{pmatrix} \tilde{\mathbf{M}}_j^{-1} \mathbf{p}_j \\ -\omega_P^2 \bar{\mathbf{M}}_j \boldsymbol{\xi}_j \end{pmatrix}}_{A} + \underbrace{\begin{pmatrix} 0 \\ -\dfrac{\partial \phi}{\partial \boldsymbol{\xi}_j} \end{pmatrix}}_{B} + \underbrace{\begin{pmatrix} 0 \\ -\gamma_{Lang}^{(j)} \mathbf{p}_j + \sigma_j \left( \tilde{\mathbf{M}}_j \right)^{1/2} \boldsymbol{\eta}_j(t) \end{pmatrix}}_{O} \quad \left( j = \overline{1, P} \right) \tag{27}
$$

with $\sigma_j = \sqrt{\dfrac{2\gamma_{Lang}^{(j)}}{\beta}}$ and each of the three parts may be solved "exactly". In case of the harmonic part (i.e., part A), the analytical solution for a time interval $\Delta t$ is

$$
\begin{aligned}
\boldsymbol{\xi}_1 &\leftarrow \boldsymbol{\xi}_1 + \tilde{\mathbf{M}}_1^{-1} \mathbf{p}_1 \Delta t \\
\begin{pmatrix} \boldsymbol{\xi}_j \\ \mathbf{p}_j \end{pmatrix} &\leftarrow \begin{pmatrix} \cos(\omega_P \Delta t)\mathbf{1} & \sin(\omega_P \Delta t)/\omega_P \, \tilde{\mathbf{M}}_j^{-1} \\ -\omega_P \sin(\omega_P \Delta t) \tilde{\mathbf{M}}_j & \cos(\omega_P \Delta t)\mathbf{1} \end{pmatrix} \begin{pmatrix} \boldsymbol{\xi}_j \\ \mathbf{p}_j \end{pmatrix} \quad \left( j = \overline{2, P} \right)
\end{aligned} \quad . \tag{28}
$$

The similar technique is often employed in MD when the harmonic system is used as the reference [22, 23]. While part B leads to

$$
\mathbf{p}_j \leftarrow \mathbf{p}_j - \dfrac{\partial \phi}{\partial \boldsymbol{\xi}_j} \Delta t \quad , \tag{29}
$$

the solution to the Ornstein-Uhlenbeck (OU) part (i.e., part O) is

$$
\mathbf{p}_j \leftarrow e^{-\gamma_{Lang}^{(j)} \Delta t} \mathbf{p}_j + \sqrt{\dfrac{1 - e^{-2\gamma_{Lang}^{(j)} \Delta t}}{\beta}} \left( \tilde{\mathbf{M}}_j \right)^{1/2} \boldsymbol{\eta}_j \quad . \tag{30}
$$

Here $\boldsymbol{\eta}_j$ is the independent Gaussian-distributed random number vector as discussed for Eq. (25).

By employing the velocity Verlet algorithm with a (white noise) Langevin thermostat for MD [24, 25], Ceriotti *et al.* constructed the PILE algorithm[17] for PIMD, which used the splitting in Eq. (27) for a time interval $\Delta t$ by the composition

$$e^{\mathcal{L}\Delta t} \approx e^{\mathcal{L}_O \Delta t/2} e^{\mathcal{L}_B \Delta t/2} e^{\mathcal{L}_A \Delta t} e^{\mathcal{L}_B \Delta t/2} e^{\mathcal{L}_O \Delta t/2} \qquad (31)$$

for the phase space propagator $e^{\mathcal{L}\Delta t}$ associated to the Langevin equation. For comparing to the new integrator that will be shortly introduced, we note it OBABO according to the order of splitting. The OBABO algorithm (or equivalently PILE) for propagating the PIMD trajectory through a time interval $\Delta t$ for Eq. (25) is

$$\mathbf{p}_j \leftarrow c_1^{(j)} \mathbf{p}_j + c_2^{(j)} \sqrt{\frac{1}{\beta}} \left(\tilde{\mathbf{M}}_j\right)^{1/2} \boldsymbol{\eta}_j \qquad \left(j = \overline{1, P}\right) \qquad (32)$$

$$\mathbf{p}_j \leftarrow \mathbf{p}_j - \frac{\partial \phi}{\partial \boldsymbol{\xi}_j} \frac{\Delta t}{2} \qquad \left(j = \overline{1, P}\right) \qquad (33)$$

$$\begin{aligned}
\boldsymbol{\xi}_1 &\leftarrow \boldsymbol{\xi}_1 + \tilde{\mathbf{M}}_1^{-1} \mathbf{p}_1 \Delta t \\
\begin{pmatrix} \boldsymbol{\xi}_j \\ \mathbf{p}_j \end{pmatrix} &\leftarrow \begin{pmatrix} \cos\left(\omega_P \Delta t\right) \mathbf{1} & \sin\left(\omega_P \Delta t\right) \big/ \omega_P \, \tilde{\mathbf{M}}_j^{-1} \\ -\omega_P \sin\left(\omega_P \Delta t\right) \tilde{\mathbf{M}}_j & \cos\left(\omega_P \Delta t\right) \mathbf{1} \end{pmatrix} \begin{pmatrix} \boldsymbol{\xi}_j \\ \mathbf{p}_j \end{pmatrix} \quad \left(j = \overline{2, P}\right)
\end{aligned} \qquad (34)$$

$$\mathbf{p}_j \leftarrow \mathbf{p}_j - \frac{\partial \phi}{\partial \boldsymbol{\xi}_j} \frac{\Delta t}{2} \qquad \left(j = \overline{1, P}\right) \qquad (35)$$

$$\mathbf{p}_j \leftarrow c_1^{(j)} \mathbf{p}_j + c_2^{(j)} \sqrt{\frac{1}{\beta}} \left(\tilde{\mathbf{M}}_j\right)^{1/2} \boldsymbol{\eta}_j \qquad \left(j = \overline{1, P}\right) \qquad (36)$$

Here the independent Gaussian-distributed random number vector $\boldsymbol{\eta}_j$ is different for each invocation of Eq. (32) or Eq. (36). The coefficients $c_1^{(j)}$ and $c_2^{(j)}$ are

$$
\begin{aligned}
c_1^{(j)} &= \exp\left[-\gamma_{Lang}^{(j)}\Delta t/2\right] \\
c_2^{(j)} &= \sqrt{1-\left(c_1^{(j)}\right)^2}
\end{aligned}
\qquad \left(j=\overline{1,P}\right) \quad,
\tag{37}
$$

respectively. While the forces in Eq. (33) are obtained from the previous time step, those in Eq. (35) at the current time step can be efficiently evaluated by the chain rule Eq. (18). The OBABO (or PILE) algorithm [Eqs. (32)-(36)] for staging PIMD has already been employed in Ref. [20].

Leimkuhler and Matthews have recently tested various algorithms for thermostatting MD with Langevin dynamics[26-28]. It is suggested that the splitting

$$
e^{\mathcal{L}\Delta t} \approx e^{\mathcal{L}_B\Delta t/2}e^{\mathcal{L}_A\Delta t/2}e^{\mathcal{L}_O\Delta t}e^{\mathcal{L}_A\Delta t/2}e^{\mathcal{L}_B\Delta t/2}
\tag{38}
$$

leads to the most efficient MD algorithm for sampling the configurational space in the high friction limit[26, 27]. When the order of splitting Eq. (38) is implemented to construct a PIMD algorithm for Eq. (27), we note it BAOAB. Such a BAOAB integrator for propagating the PIMD trajectory through a time interval $\Delta t$ for Eq. (25) is

$$
\mathbf{p}_j \leftarrow \mathbf{p}_j - \frac{\partial\phi}{\partial\boldsymbol{\xi}_j}\frac{\Delta t}{2} \qquad \left(j=\overline{1,P}\right)
\tag{39}
$$

$$
\begin{aligned}
\boldsymbol{\xi}_1 &\leftarrow \boldsymbol{\xi}_1 + \tilde{\mathbf{M}}_1^{-1}\mathbf{p}_1\frac{\Delta t}{2} \\
\begin{pmatrix}\boldsymbol{\xi}_j \\ \mathbf{p}_j\end{pmatrix} &\leftarrow \begin{pmatrix} \cos\left(\omega_P\Delta t/2\right)\mathbf{1} & \sin\left(\omega_P\Delta t/2\right)/\omega_P\,\tilde{\mathbf{M}}_j^{-1} \\ -\omega_P\sin\left(\omega_P\Delta t/2\right)\tilde{\mathbf{M}}_j & \cos\left(\omega_P\Delta t/2\right)\mathbf{1}\end{pmatrix}\begin{pmatrix}\boldsymbol{\xi}_j \\ \mathbf{p}_j\end{pmatrix} \quad \left(j=\overline{2,P}\right)
\end{aligned}
\tag{40}
$$

$$\mathbf{p}_j \leftarrow \tilde{c}_1^{(j)} \mathbf{p}_j + \tilde{c}_2^{(j)} \sqrt{\frac{1}{\beta}} \left(\tilde{\mathbf{M}}_j\right)^{1/2} \boldsymbol{\eta}_j \qquad \left(j = \overline{1,P}\right) \tag{41}$$

$$\boldsymbol{\xi}_1 \leftarrow \boldsymbol{\xi}_1 + \tilde{\mathbf{M}}_1^{-1} \mathbf{p}_1 \frac{\Delta t}{2}$$
$$\begin{pmatrix} \boldsymbol{\xi}_j \\ \mathbf{p}_j \end{pmatrix} \leftarrow \begin{pmatrix} \cos\left(\omega_P \Delta t/2\right) \mathbf{1} & \sin\left(\omega_P \Delta t/2\right)/\omega_P \, \tilde{\mathbf{M}}_j^{-1} \\ -\omega_P \sin\left(\omega_P \Delta t/2\right) \tilde{\mathbf{M}}_j & \cos\left(\omega_P \Delta t/2\right) \mathbf{1} \end{pmatrix} \begin{pmatrix} \boldsymbol{\xi}_j \\ \mathbf{p}_j \end{pmatrix} \quad \left(j = \overline{2,P}\right) \tag{42}$$

$$\mathbf{p}_j \leftarrow \mathbf{p}_j - \frac{\partial \phi}{\partial \boldsymbol{\xi}_j} \frac{\Delta t}{2} \qquad \left(j = \overline{1,P}\right) \quad , \tag{43}$$

where the independent Gaussian-distributed random number vector $\boldsymbol{\eta}_j$ is different for each invocation of Eq. (41), and the coefficients are

$$\tilde{c}_1^{(j)} = \exp\left[-\gamma_{Lang}^{(j)} \Delta t\right]$$
$$\tilde{c}_2^{(j)} = \sqrt{1 - \left(\tilde{c}_1^{(j)}\right)^2} \qquad \left(j = \overline{1,P}\right) \quad . \tag{44}$$

The implementation of the BAOAB algorithm [Eqs. (39)-(43)] is very simple, which has already been done in our earlier work[21, 29].

## 4. Accuracy of the PIMD integrators

BAOAB and OBABO approach each other in the limit $\Delta t \to 0$. Both BAOAB and OBABO exploit an analytic knowledge of path integral staging mode frequencies in the free particle limit [Eq. (26)]. It is trivial to verify that either of BAOAB and OBABO is exact in the free particle limit.

Eq. (27) can be expressed in a more compact form as

$$\begin{pmatrix} \dot{\boldsymbol{\xi}} \\ \dot{\mathbf{p}} \end{pmatrix} = \underbrace{\begin{pmatrix} \tilde{\mathbf{M}}^{-1}\,\mathbf{p} \\ -\omega_P^2 \bar{\mathbf{M}} \boldsymbol{\xi} \end{pmatrix}}_{A} + \underbrace{\begin{pmatrix} 0 \\ -\dfrac{\partial \phi}{\partial \boldsymbol{\xi}} \end{pmatrix}}_{B} + \underbrace{\begin{pmatrix} 0 \\ -\boldsymbol{\gamma}_{Lang}\mathbf{p} + \boldsymbol{\sigma}\,\tilde{\mathbf{M}}^{1/2}\boldsymbol{\eta}(t) \end{pmatrix}}_{O} \quad . \tag{45}$$

Here $\boldsymbol{\xi} = \begin{pmatrix} \boldsymbol{\xi}_1 \\ \vdots \\ \boldsymbol{\xi}_P \end{pmatrix}$, $\mathbf{p} = \begin{pmatrix} \mathbf{p}_1 \\ \vdots \\ \mathbf{p}_P \end{pmatrix}$, $\bar{\mathbf{M}} = \begin{pmatrix} \bar{\mathbf{M}}_1 & & \\ & \ddots & \\ & & \bar{\mathbf{M}}_P \end{pmatrix}$, $\tilde{\mathbf{M}} = \begin{pmatrix} \tilde{\mathbf{M}}_1 & & \\ & \ddots & \\ & & \tilde{\mathbf{M}}_P \end{pmatrix}$, $\boldsymbol{\eta}(t) = \begin{pmatrix} \boldsymbol{\eta}_1(t) \\ \vdots \\ \boldsymbol{\eta}_P(t) \end{pmatrix}$,

$\boldsymbol{\gamma}_{Lang} = \begin{pmatrix} \gamma_{Lang}^{(1)}\cdot\mathbf{1}_{N\times N} & & \\ & \ddots & \\ & & \gamma_{Lang}^{(P)}\cdot\mathbf{1}_{N\times N} \end{pmatrix}$, and $\boldsymbol{\sigma} = \begin{pmatrix} \sigma_1\cdot\mathbf{1}_{N\times N} & & \\ & \ddots & \\ & & \sigma_P\cdot\mathbf{1}_{N\times N} \end{pmatrix}$.

Eq. (20) then becomes

$$\begin{aligned} H_{eff}\left(\boldsymbol{\xi};\mathbf{p}\right) &= \frac{1}{2}\mathbf{p}^T\tilde{\mathbf{M}}^{-1}\mathbf{p} + \frac{1}{2}\omega_P^2\boldsymbol{\xi}^T\bar{\mathbf{M}}\boldsymbol{\xi} + \phi(\boldsymbol{\xi}) \\ &= \frac{1}{2}\mathbf{p}^T\tilde{\mathbf{M}}^{-1}\mathbf{p} + U_{eff}\left(\boldsymbol{\xi}\right) \end{aligned} \quad . \tag{46}$$

The density evolves according to the Fokker-Planck or forward Kolmogorov equation

$$\frac{\partial \rho}{\partial t} = \mathcal{L}\rho \quad , \tag{47}$$

where the relevant Kolmogorov operator $\mathcal{L}$ is defined by

$$\mathcal{L} = \mathcal{L}_{LD} = \mathcal{L}_A + \mathcal{L}_B + \mathcal{L}_O \tag{48}$$

with

$$\mathcal{L}_A\rho = -\mathbf{p}^T\tilde{\mathbf{M}}^{-1}\frac{\partial}{\partial \boldsymbol{\xi}}\rho + \omega_P^2\boldsymbol{\xi}^T\bar{\mathbf{M}}\frac{\partial}{\partial \mathbf{p}}\rho \quad , \tag{49}$$

$$\mathcal{L}_B \rho = \left( \frac{\partial \phi}{\partial \boldsymbol{\xi}} \right)^T \frac{\partial}{\partial \mathbf{p}} \rho \quad , \tag{50}$$

$$\mathcal{L}_O \rho = \frac{\partial}{\partial \mathbf{p}} \cdot \left( \boldsymbol{\gamma}_{Lang} \mathbf{p} \rho \right) + \frac{1}{2} \frac{\partial}{\partial \mathbf{p}} \cdot \left( \boldsymbol{\sigma}^2 \tilde{\mathbf{M}} \frac{\partial \rho}{\partial \mathbf{p}} \right) \quad . \tag{51}$$

It is straightforward to verify that

$$\rho_{eq}(\boldsymbol{\xi}; \mathbf{p}) = Z_{eff}^{-1} \exp\left[ -\beta H_{eff}(\boldsymbol{\xi}; \mathbf{p}) \right] \tag{52}$$

is a steady state of Eq. (47). Here $Z_{eff}$ is the normalization constant of the density distribution

$$Z_{eff} = \int d\boldsymbol{\xi} d\mathbf{p} \exp\left\{ -\beta H_{eff}(\boldsymbol{\xi}; \mathbf{p}) \right\} = \int \left( \prod_{j=1}^{P} d\boldsymbol{\xi}_j d\mathbf{p}_j \right) \exp\left\{ -\beta H_{eff}(\boldsymbol{\xi}_1, \cdots, \boldsymbol{\xi}_P; \mathbf{p}_1, \cdots, \mathbf{p}_P) \right\}.$$

The exact phase space propagator for a time interval $\Delta t$ for Eq. (47) is $e^{\mathcal{L}_{LD} \Delta t}$. When the BAOAB integrator [Eq. (38)] is employed, the "approximate" phase space propagator is

$$e^{\mathcal{L}_B \Delta t/2} e^{\mathcal{L}_A \Delta t/2} e^{\mathcal{L}_O \Delta t} e^{\mathcal{L}_A \Delta t/2} e^{\mathcal{L}_B \Delta t/2} = e^{\mathcal{L}_{BAOAB} \Delta t} \quad . \tag{53}$$

Using the Baker–Campbell–Hausdorff formula to expand the LHS of Eq. (53), one then obtains

$$\begin{aligned}
\mathcal{L}_{BAOAB} &= \mathcal{L}_A + \mathcal{L}_B + \mathcal{L}_O + \frac{1}{24} \Big\{ 2\big[ \mathcal{L}_O, [\mathcal{L}_O, \mathcal{L}_A + \mathcal{L}_B] \big] + 2\big[ \mathcal{L}_A, [\mathcal{L}_A, \mathcal{L}_B] \big] \\
&\quad + 2\big[ \mathcal{L}_O, [\mathcal{L}_A, \mathcal{L}_B] \big] + 2\big[ \mathcal{L}_A, [\mathcal{L}_O, \mathcal{L}_B] \big] - \big[ \mathcal{L}_A, [\mathcal{L}_A, \mathcal{L}_O] \big] \\
&\quad - \big[ \mathcal{L}_B, [\mathcal{L}_B, \mathcal{L}_O] \big] - \big[ \mathcal{L}_B, [\mathcal{L}_B, \mathcal{L}_A] \big] \Big\} \Delta t^2 + O(\Delta t^4) \\
&= \mathcal{L}_{LD} + \mathcal{L}_2^{BAOAB} \Delta t^2 + O(\Delta t^4)
\end{aligned} \tag{54}$$

It is straightforward to show that

$$\mathcal{L}_2^{BAOAB} \rho_{eq} = \rho_{eq} \left[ \frac{1}{4} \left( \frac{\partial}{\partial \boldsymbol{\xi}} \cdot \left( \boldsymbol{\gamma}_{Lang} \tilde{\mathbf{M}}^{-1} \frac{\partial \phi}{\partial \boldsymbol{\xi}} \right) - \beta \mathbf{p}^T \boldsymbol{\gamma}_{Lang} \tilde{\mathbf{M}}^{-1} \phi'' \tilde{\mathbf{M}}^{-1} \mathbf{p} \right) \right.$$

$$+ \frac{\beta}{4} \mathbf{p}^T \tilde{\mathbf{M}}^{-1} \phi'' \tilde{\mathbf{M}}^{-1} \frac{\partial \phi}{\partial \boldsymbol{\xi}} - \frac{\beta}{12} \mathbf{p}^T \tilde{\mathbf{M}}^{-1} \frac{\partial}{\partial \boldsymbol{\xi}} \left( \mathbf{p}^T \tilde{\mathbf{M}}^{-1} \phi'' \tilde{\mathbf{M}}^{-1} \mathbf{p} \right) \quad , \qquad (55)$$

$$\left. + \frac{\beta}{12} \omega_P^2 \left( \mathbf{p}^T \tilde{\mathbf{M}}^{-1} \bar{\mathbf{M}} \tilde{\mathbf{M}}^{-1} \frac{\partial \phi}{\partial \boldsymbol{\xi}} + 3 \mathbf{p}^T \tilde{\mathbf{M}}^{-1} \phi'' \bar{\mathbf{M}} \tilde{\mathbf{M}}^{-1} \boldsymbol{\xi} \right) \right]$$

where $\phi'' = \dfrac{\partial^2 \phi}{\partial \boldsymbol{\xi}^2}$ is a Hessian matrix.

Consider the steady state $\rho^{BAOAB}$ for the relevant Kolmogorov operator $\mathcal{L}_{BAOAB}$, which satisfies

$$\frac{\partial \rho^{BAOAB}}{\partial t} = \mathcal{L}_{BAOAB} \rho^{BAOAB} = 0 \qquad . \qquad (56)$$

Assume that $\rho^{BAOAB}$ takes the form

$$\rho^{BAOAB} = \rho_{eq} \left[ 1 - \beta \omega_P^2 \Delta t^2 f_2^{BAOAB} + O(\omega_P^4 \Delta t^4) \right] \quad . \qquad (57)$$

Substituting Eq. (54) and Eq. (57) into Eq. (56) leads to

$$\mathcal{L}_{LD}(\rho_{eq} f_2^{BAOAB}) = \frac{1}{\beta \omega_P^2} \mathcal{L}_2^{BAOAB} \rho_{eq} \qquad . \qquad (58)$$

When Eq. (26) is used as the optimum friction coefficients, $\boldsymbol{\gamma}_{Lang}$ is then expressed as $\boldsymbol{\gamma}_{Lang} = \omega_P \boldsymbol{\gamma}_1$. Since the number of beads $P$ is often large, $\varepsilon = 1/\omega_P$ is small. $f_2^{BAOAB}$ can be expressed as

$$f_2^{BAOAB} = f_{2,0}^{BAOAB} + f_{2,1}^{BAOAB} \varepsilon + f_{2,2}^{BAOAB} \varepsilon^2 + O(\varepsilon^3) \quad . \qquad (59)$$

Substituting Eq. (59) into Eq. (58) and then dividing both sides by $\omega_p^2$, one finds

$$
\begin{aligned}
&\left(\mathcal{L}_0 + \varepsilon \mathcal{L}_1 + \varepsilon^2 \mathcal{L}_2\right)\left[f_{2,0}^{BAOAB} + f_{2,1}^{BAOAB}\varepsilon + f_{2,2}^{BAOAB}\varepsilon^2 + O(\varepsilon^3)\right] \\
&= g_2^{BAOAB}\varepsilon^2 + g_3^{BAOAB}\varepsilon^3 + g_4^{BAOAB}\varepsilon^4
\end{aligned} \quad,
\tag{60}
$$

where

$$
\mathcal{L}_0 \rho = \boldsymbol{\xi}^T \bar{\mathbf{M}} \frac{\partial}{\partial \mathbf{p}} \rho \quad,
\tag{61}
$$

$$
\mathcal{L}_1 \rho = \frac{1}{\beta} \frac{\partial}{\partial \mathbf{p}} \cdot \left(\boldsymbol{\gamma}_1 \tilde{\mathbf{M}} \frac{\partial \rho}{\partial \mathbf{p}}\right) - \mathbf{p}^T \boldsymbol{\gamma}_1 \frac{\partial \rho}{\partial \mathbf{p}} \quad,
\tag{62}
$$

$$
\mathcal{L}_2 \rho = \left(\frac{\partial \phi}{\partial \boldsymbol{\xi}}\right)^T \frac{\partial \rho}{\partial \mathbf{p}} - \mathbf{p}^T \tilde{\mathbf{M}}^{-1} \frac{\partial \rho}{\partial \boldsymbol{\xi}} \quad,
\tag{63}
$$

$$
g_2^{BAOAB} = \frac{1}{12}\left(\mathbf{p}^T \tilde{\mathbf{M}}^{-1} \bar{\mathbf{M}} \tilde{\mathbf{M}}^{-1} \frac{\partial \phi}{\partial \boldsymbol{\xi}} + 3\mathbf{p}^T \tilde{\mathbf{M}}^{-1} \boldsymbol{\phi}'' \bar{\mathbf{M}} \tilde{\mathbf{M}}^{-1} \boldsymbol{\xi}\right) \quad,
\tag{64}
$$

$$
g_3^{BAOAB} = \frac{1}{4}\left(\frac{1}{\beta} \frac{\partial}{\partial \boldsymbol{\xi}} \cdot \left(\boldsymbol{\gamma}_1 \tilde{\mathbf{M}}^{-1} \frac{\partial \phi}{\partial \boldsymbol{\xi}}\right) - \mathbf{p}^T \boldsymbol{\gamma}_1 \tilde{\mathbf{M}}^{-1} \boldsymbol{\phi}'' \tilde{\mathbf{M}}^{-1} \mathbf{p}\right) \quad,
\tag{65}
$$

$$
g_4^{BAOAB} = \frac{1}{4} \mathbf{p}^T \tilde{\mathbf{M}}^{-1} \boldsymbol{\phi}'' \tilde{\mathbf{M}}^{-1} \frac{\partial \phi}{\partial \boldsymbol{\xi}} - \frac{1}{12} \mathbf{p}^T \tilde{\mathbf{M}}^{-1} \frac{\partial}{\partial \boldsymbol{\xi}}\left(\mathbf{p}^T \tilde{\mathbf{M}}^{-1} \boldsymbol{\phi}'' \tilde{\mathbf{M}}^{-1} \mathbf{p}\right) \quad.
\tag{66}
$$

Equating powers of $\varepsilon$ in Eq. (60), one obtains

$$
\begin{aligned}
&\mathcal{L}_0 f_{2,0}^{BAOAB} = 0 \quad, \\
&\mathcal{L}_1 f_{2,0}^{BAOAB} + \mathcal{L}_0 f_{2,1}^{BAOAB} = 0 \quad, \\
&\mathcal{L}_0 f_{2,2}^{BAOAB} + \mathcal{L}_1 f_{2,1}^{BAOAB} + \mathcal{L}_2 f_{2,0}^{BAOAB} = g_2^{BAOAB} \quad, \\
&\cdots
\end{aligned}
\tag{67}
$$

Truncating at the $2^{nd}$ order of $\varepsilon$, one finds a solution of Eq. (67)

$$
\begin{aligned}
f_{2,0}^{BAOAB} &= -\frac{1}{12}\phi + G_0 \\
f_{2,1}^{BAOAB} &= \frac{1}{12}\mathbf{p}^T\boldsymbol{\gamma}_1^{-1}\tilde{\mathbf{M}}^{-1}(\mathbf{1}-\bar{\mathbf{M}}\tilde{\mathbf{M}}^{-1})\frac{\partial\phi}{\partial\boldsymbol{\xi}} - G_1(\boldsymbol{\xi}) \\
f_{2,2}^{BAOAB} &= \frac{1}{8}\left[\mathbf{p}^T\tilde{\mathbf{M}}^{-1}\phi''\tilde{\mathbf{M}}^{-1}\mathbf{p} - \frac{1}{\beta}\frac{\partial}{\partial\boldsymbol{\xi}}\cdot\left(\tilde{\mathbf{M}}^{-1}\frac{\partial\phi}{\partial\boldsymbol{\xi}}\right)\right] - G_2(\boldsymbol{\xi})
\end{aligned}
\qquad (68)
$$

Here $G_0$ is a constant, $G_1(\boldsymbol{\xi})$ and $G_2(\boldsymbol{\xi})$ are functions to be determined by the equations for the $3^{rd}$ and higher orders of $\varepsilon$ in Eq. (67). (In the free particle limit, $G_0$, $G_1(\boldsymbol{\xi})$, and $G_2(\boldsymbol{\xi})$ all approach zero.)

Note that only the configurational distribution of PIMD is useful, i.e.,

$$
\rho_{eq}^{config}(\boldsymbol{\xi}) = \int d\mathbf{p}\,\rho_{eq}(\boldsymbol{\xi};\mathbf{p}) = \frac{1}{Z^{config}}\exp\left[-\beta U_{eff}(\boldsymbol{\xi})\right]
\qquad (69)
$$

with a new normalization coefficient $Z^{config}$. Integration of $\rho^{BAOAB}$ [Eq. (57)] over $\mathbf{p}$ produces

$$
\begin{aligned}
\rho_{BAOAB}^{config}(\boldsymbol{\xi}) = \rho_{eq}^{config}(\boldsymbol{\xi})\Big\{&1 + \beta\omega_P^2\Delta t^2\big[(\phi/12 - G_0) \\
&+ \varepsilon G_1(\boldsymbol{\xi}) + \varepsilon^2 G_2(\boldsymbol{\xi}) + O(\varepsilon^3)\big] + O(\omega_P^4\Delta t^4)\Big\}
\end{aligned}
\qquad (70)
$$

It is difficult to analyze Eq. (70) for general systems. Consider a harmonic oscillator

$$
V(\mathbf{x}) = \frac{1}{2}(\mathbf{x} - \mathbf{x}_{eq})^T\mathbf{A}(\mathbf{x} - \mathbf{x}_{eq}) \quad,
\qquad (71)
$$

where $\mathbf{A}$ is a symmetric positive-definite matrix. Eq. (17) then becomes

$$\phi = \frac{1}{2}(\xi - \xi_{eq})^T \mathbf{K}(\xi - \xi_{eq}) \quad . \tag{72}$$

Here $\xi = \mathbf{S}\begin{pmatrix} \mathbf{x}_1 \\ \vdots \\ \mathbf{x}_P \end{pmatrix}$,

$$\xi_{eq} = \mathbf{S}\begin{pmatrix} \mathbf{x}_{eq} \\ \vdots \\ \mathbf{x}_{eq} \end{pmatrix} = \begin{pmatrix} \mathbf{x}_{eq} \\ 0 \\ \vdots \\ 0 \end{pmatrix} \quad , \tag{73}$$

and the symmetric positive-definite matrix

$$\mathbf{K} = \frac{1}{P}(\mathbf{S}^{-1})^T \begin{pmatrix} \mathbf{A} & & \\ & \ddots & \\ & & \mathbf{A} \end{pmatrix}(\mathbf{S}^{-1}) \quad , \tag{74}$$

with the staging transformation matrix

$$\mathbf{S} = \begin{pmatrix} \mathbf{1} & 0 & 0 & \cdots & 0 \\ -\dfrac{\mathbf{1}}{\mathbf{2}} & \mathbf{1} & -\dfrac{\mathbf{1}}{\mathbf{2}} & \ddots & \vdots \\ \vdots & 0 & \ddots & \ddots & 0 \\ -\dfrac{\mathbf{1}}{\mathbf{P-1}} & \vdots & 0 & \mathbf{1} & -\dfrac{\mathbf{P-2}}{\mathbf{P-1}} \\ -\mathbf{1} & 0 & \cdots & 0 & \mathbf{1} \end{pmatrix} \tag{75}$$

and its inverse

$$\mathbf{S}^{-1} = \begin{pmatrix} \mathbf{1} & 0 & 0 & 0 & \cdots & 0 \\ \mathbf{1} & \mathbf{1} & \dfrac{\mathbf{1}}{\mathbf{2}} & \dfrac{\mathbf{1}}{\mathbf{3}} & \cdots & \dfrac{\mathbf{1}}{\mathbf{P-1}} \\ \mathbf{1} & 0 & \mathbf{1} & \dfrac{\mathbf{2}}{\mathbf{3}} & \cdots & \dfrac{\mathbf{2}}{\mathbf{P-1}} \\ \vdots & \vdots & \ddots & \ddots & & \vdots \\ \mathbf{1} & 0 & \cdots & 0 & \mathbf{1} & \dfrac{\mathbf{P-2}}{\mathbf{P-1}} \\ \mathbf{1} & 0 & \cdots & & 0 & \mathbf{1} \end{pmatrix} \qquad . \tag{76}$$

Each nonzero number in Eq. (75) or (76) represents a diagonal $N \times N$ matrix. Because Eq.(16) and Eq. (73) lead to

$$\bar{\mathbf{M}}\boldsymbol{\xi}_{eq} = \begin{pmatrix} 0 & & & \\ & \bar{\mathbf{M}}_2 & & \\ & & \ddots & \\ & & & \bar{\mathbf{M}}_P \end{pmatrix} \begin{pmatrix} \mathbf{x}_{eq} \\ 0 \\ \vdots \\ 0 \end{pmatrix} = 0 \quad , \tag{77}$$

Eq. (21) then becomes

$$U_{eff}\left(\boldsymbol{\xi}\right) = \frac{1}{2}(\boldsymbol{\xi} - \boldsymbol{\xi}_{eq})^T \boldsymbol{\Omega}(\boldsymbol{\xi} - \boldsymbol{\xi}_{eq}) \quad , \tag{78}$$

where

$$\boldsymbol{\Omega} = \omega_P^2 \bar{\mathbf{M}} + \mathbf{K} \tag{79}$$

is a symmetric positive-definite matrix.

It is straightforward to verify that Eq. (70) becomes

$$\rho_{BAOAB}^{config}(\xi) = \rho_{eq}^{config}(\xi)\left\{1 + \beta\omega_P^2\Delta t^2\left[\frac{1}{24}(\xi - \xi_{eq})^T \mathbf{K}(\xi - \xi_{eq}) - \frac{1}{24\beta}\mathrm{Tr}\left(\mathbf{K}\mathbf{\Omega}^{-1}\right)\right.\right.$$
$$\left.\left. + O(\varepsilon)\right] + O\left(\omega_P^4\Delta t^4\right)\right\} \tag{80}$$

while truncating at the 0-th order of $\varepsilon$ for the term associated with $\omega_P^2\Delta t^2$. Eq. (80) is a normalized density distribution.

Similarly, the normalized configuration distribution given by OBABO is

$$\rho_{OBABO}^{config}(\xi) = \rho_{eq}^{config}(\xi)\left\{1 + \beta\omega_P^2\Delta t^2\left[\left(\frac{1}{8}\xi^T\bar{\mathbf{M}}\tilde{\mathbf{M}}^{-1}\frac{\partial\phi}{\partial\xi} - \frac{1}{6}\phi + \tilde{G}_0\right)\right.\right.$$
$$+ \varepsilon\tilde{G}_1(\xi) + \varepsilon^2\left(\frac{1}{8}\left(\frac{\partial\phi}{\partial\xi}\right)^T\tilde{\mathbf{M}}^{-1}\frac{\partial\phi}{\partial\xi} - \frac{1}{8}\frac{1}{\beta}\frac{\partial}{\partial\xi}\cdot\left(\tilde{\mathbf{M}}^{-1}\frac{\partial\phi}{\partial\xi}\right) + \tilde{G}_2(\xi)\right)$$
$$\left.\left. + O(\varepsilon^3)\right] + O\left(\omega_P^4\Delta t^4\right)\right\} \tag{81}$$

It is easy to show that Eq. (81) for the harmonic system [Eq. (71) or Eq. (72)] becomes

$$\rho_{OBABO}^{config}(\xi) = \rho_{eq}^{config}(\xi)\left\{1 + \beta\omega_P^2\Delta t^2\left[(\xi - \xi_{eq})^T\left(\frac{1}{4}\bar{\mathbf{M}}\tilde{\mathbf{M}}^{-1} - \frac{1}{12}\mathbf{1}\right)\mathbf{K}(\xi - \xi_{eq})\right.\right.$$
$$\left.\left. - \frac{1}{\beta}\mathrm{Tr}\left(\left(\frac{1}{4}\bar{\mathbf{M}}\tilde{\mathbf{M}}^{-1} - \frac{1}{12}\mathbf{1}\right)\mathbf{K}\mathbf{\Omega}^{-1}\right) + O(\varepsilon)\right] + O\left(\omega_P^4\Delta t^4\right)\right\} \tag{82}$$

while truncating at the 0-th order of $\varepsilon$ for the term associated with $\omega_P^2\Delta t^2$.

Comparing Eq. (82) to Eq. (80), the error (of the configurational distribution) produced by OBABO is increased by approximately a factor of 4 of that by BAOAB for the harmonic system. It is trivial to extend the BAOAB algorithm and the error analysis to normal mode PIMD, which leads to the similar conclusion (as shown in Supplemental Material[18]).

**III. Results and discussions**

We apply the two PIMD integrators to three benchmark realistic molecular systems—the water molecule ($H_2O$) with an accurate PES[30], liquid *para*-hydrogen with the Silvera-Goldman (SG) potential[31], and liquid water with an *ab initio* based flexible, polarizable force field[32]. Both the primitive and virial estimators [Eqs. (7) and (8)] are employed to calculate the average kinetic energy. The average potential energy is also computed. These thermodynamic properties are collected as a function of the time interval of the PIMD simulation. In practice, when the number of beads $P$ is large enough and the time interval $\Delta t$ is small enough, both BAOAB and OBABO lead to the same converged results. So do the primitive and virial estimators.

**1. The $H_2O$ molecule**

We first apply the two integrators to simulate $H_2O$ with the accurate PES developed by Partridge and Schwenke from extensive *ab initio* calculations and experimental data[30]. As the explicit form of the PES is available, that of the force can be expressed. $P = 640$ path integral beads are used in PIMD for $T = 100\,\text{K}$, while $P = 256$ for $T = 300\,\text{K}$. After equilibrating the molecular system, 32 PIMD trajectories with each propagated up to ~4.84 ns are used for estimating thermodynamic properties. While the time interval for PIMD for 300K ranges from ~0.024 fs to ~0.39 fs (1 − 16 au), that for 100K is from ~0.012 fs to ~0.32 fs (0.5 − 13 au).

Fig. 1 demonstrates the results for the thermodynamic properties using different time intervals of PIMD for $H_2O$ at $T = 300\,\text{K}$, while Fig. 2 does so for $T = 100\,\text{K}$. The performance of BAOAB and that of OBABO are examined. Figs 1 and 2 show that BAOAB and OBABO approach the same results as the time interval is decreased. This agrees with the fact that both integrators are in principle equivalent as the time interval approaches zero. Figs 1a and 2a also show that the primitive estimator for the kinetic energy agrees well with the virial estimator when the time interval is small. When the number of path integral beads $P$ is reasonably large, the primitive estimator in principle approaches the virial one as the time interval $\Delta t$ of PIMD approaches zero. The difference between the results of the primitive and virial estimators $\Delta E_{kin}$ is then a reasonable quantity for measuring the behavior of the integrator for PIMD. The fully converged result for either $\left\langle \hat{\mathbf{p}}^T \mathbf{M}^{-1} \hat{\mathbf{p}} \right\rangle / \left( 2 N_{atom} k_B \right)$ or $\left\langle V(\hat{\mathbf{x}}) \right\rangle / \left( N_{atom} k_B \right)$ is obtained at $\Delta t = 1$ au .

More importantly, Figs 1 and 2 demonstrate that BAOAB is more accurate and robust than OBABO while the time interval $\Delta t$ increases. While the absolute deviation of the average potential energy per atom $\left\langle V(\hat{\mathbf{x}}) \right\rangle / \left( N_{atom} k_B \right)$ from the converged result for OBABO at $T = 300$ K is ~1K for $\Delta t = 2$ au and ~37K for $\Delta t = 10$ au (Fig. 1c), that at $T = 100$ K is ~0.8K and ~32K, respectively (Fig. 2c). For comparison, the same property for BAOAB at $T = 300$ K is only ~0.1K for $\Delta t = 2$ au and ~1K for $\Delta t = 10$ au (Fig. 1c), that at $T = 100$ K is only ~0.2 K and ~0.3K, respectively (Fig. 2c). As the time interval increases, the error of deviation from the converged result of OBABO is about an order of magnitude (or more) larger than that of BAOAB. The trend is similar for the average kinetic energy per atom $\left\langle \hat{\mathbf{p}}^T \mathbf{M}^{-1} \hat{\mathbf{p}} \right\rangle / \left( 2 N_{atom} k_B \right)$ (by either the primitive or virial estimator) or for the difference between the two estimators $\Delta E_{kin}$ as suggested by Figs 1a-b and 2a-b.

## 2. Liquid *para*-hydrogen

Liquid *para*-hydrogen is usually described by the Silvera-Goldman (SG) model[31], an isotropic pair potential in which the *para*-hydrogen molecule is treated as a sphere particle because the temperature of liquid *para*-hydrogen is much too low for any rotational state other than J=0 to be populated. Liquid *para*-hydrogen has served as a benchmark system to test quantum methods. We apply the two PIMD integrators to two state points $T = 25\,\mathrm{K}, \upsilon = 31.7\,\mathrm{cm^3mol^{-1}}$ and $T = 14\,\mathrm{K}, \upsilon = 25.6\,\mathrm{cm^3mol^{-1}}$ under nearly zero external pressure. PIMD simulations are carried out for a system of 125 *para*-hydrogen molecules in a box with periodic boundary conditions applied using the minimum image convention. $P = 48$ path integral beads are employed in PIMD for $T = 25\,\mathrm{K}$, while $P = 96$ for $T = 14\,\mathrm{K}$. After equilibrating the system, 16 PIMD trajectories with each propagated up to ~24.2 ns are used for estimating thermodynamic properties. While the time interval for PIMD is from ~1.2 fs to ~12.1 fs (50~500 au) for $T = 25\,\mathrm{K}$, that for $T = 14\,\mathrm{K}$ ranges from ~1.2 fs to ~10.2 fs (50~420 au).

Figs 3 and 4 depict comparison of the performance of OBABO to that of BAOAB for simulating liquid *para*-hydrogen at $T = 25\,\mathrm{K}$ and $T = 14\,\mathrm{K}$, respectively. As shown in Fig. 3a-b for $T = 25\,\mathrm{K}$, while the difference between the primitive and virial estimators $\Delta E_{kin}$ of OBABO is ~0.25 K for $\Delta t = 200\,\mathrm{au}$ and ~1.6 K for $\Delta t = 450\,\mathrm{au}$, that of BAOAB is only ~0.012 K and ~0.13 K, respectively. Even when OBABO fails at $\Delta t = 500\,\mathrm{au}$, BAOAB still leads to only ~0.18 K for the difference $\Delta E_{kin}$. Similarly, as presented from Fig. 4a-b for $T = 14\,\mathrm{K}$, while the difference $\Delta E_{kin}$ for OBABO is ~0.11 K for $\Delta t = 100\,\mathrm{au}$ and ~1.7 K for $\Delta t = 370\,\mathrm{au}$, that for BAOAB is less than 0.01 K and ~0.09 K, respectively. OBABO fails when the time interval is larger than 390 au, however, BAOAB still produces ~0.11 K for the difference $\Delta E_{kin}$ for $\Delta t = 420\,\mathrm{au}$. Even when the time interval of BAOAB is 4~6 times of that of OBABO, the two PIMD integrators lead to comparable accuracy. This is also suggested by Fig. 3c and Fig. 4c where the average potential energy per molecule $\langle V(\hat{\mathbf{x}})\rangle/(N_{mol}k_B)$ is examined.

## 3. Liquid water

For the simulation of liquid water, we employ the TTM3-F—the *ab initio* based flexible, polarizable Thole-type model for water clusters and liquid water of Fanourgakis and Xantheas[32]. It approximates the Born-Oppenheimer potential energy surface based on the parameterization which reproduces the binding energies and harmonic vibrational spectra of small water clusters up to $(H_2O)_{20}$ given by the second order Møller-Plesset (MP2) electronic structure theory. The TTM3-F model is able to produce good results for static equilibrium structural properties of liquid water with PIMD simulations[32]. PIMD simulations are carried out at $T$=300 K with the liquid density $\rho_l = 0.997\,\text{g}\cdot\text{cm}^{-3}$ for a system of 125 water molecules in a box with periodic boundary conditions applied using the minimum image convention. $P = 72$ path integral beads are employed in the simulation. After equilibrating the system, 16 PIMD trajectories with each propagated up to ~50 ps are used for estimating thermodynamic properties. The time interval for PIMD is from 0.05 fs to 0.8 fs.

As presented in Fig. 5a, when the time interval is $\Delta t = 0.05\,\text{fs}$, both BAOAB and OBABO lead to the converged result for either the primitive or the virial estimators for the property $\langle\hat{\mathbf{p}}^T\mathbf{M}^{-1}\hat{\mathbf{p}}\rangle/(2N_{atom}k_B)$. The difference $\Delta E_{kin}$ between the results of the primitive and virial estimators is nearly zero for $\Delta t = 0.05\,\text{fs}$. While the difference $\Delta E_{kin}$ for the BAOAB integrator is only ~1.7 K for $\Delta t = 0.6\,\text{fs}$, that for the OBABO is ~2.5 K for $\Delta t = 0.15\,\text{fs}$ and ~34 K for $\Delta t = 0.6\,\text{fs}$. As the time interval increases, OBABO produces ~50 K for the difference $\Delta E_{kin}$ for

$\Delta t = 0.7\,\text{fs}$ and fails for $\Delta t = 0.8\,\text{fs}$. For comparison, the BAOAB result is only ~ 5 K for

$\Delta t = 0.8\,\text{fs}$. Fig. 5b demonstrates that the average potential energy per atom $\langle V(\hat{\mathbf{x}}) \rangle / (N_{atom} k_B)$ obtained by BAOAB agrees well with that by OBABO within the statistical error bar for the time interval $\Delta t = 0.05\,\text{fs}$, which can be considered as the converged result. While the deviation from the converged result is ~6.5 K for $\Delta t = 0.15\,\text{fs}$ and ~53 K for $\Delta t = 0.6\,\text{fs}$ for the OBABO integrator, the BAOAB result for $\Delta t = 0.6\,\text{fs}$ is still close to the converged result (within the statistical error

bar). Even when OBABO leads to ~90 K for the deviation (from the converged result) for

$\Delta t = 0.7\,\text{fs}$ and fails for $\Delta t = 0.8\,\text{fs}$, BAOAB produces only ~11 K for the deviation (from the

converged result) for $\Delta t = 0.8\,\text{fs}$. In summary, the time interval can be increased by a factor of

4~6 in BAOAB for achieving the same accuracy as OBABO does.

Interestingly, in terms of accuracy as a function of the time interval, the primitive estimator

Eq. (7) behaves consistently better than the virial one Eq. (8) when $P$ is fixed for almost all the

cases (Figs 1-5a) studied in the present paper. As shown in Fig. 4a for liquid *para*-hydrogen at

$T = 14\,\text{K}$, the absolute value of the deviation from the converged result for the primitive estimator

at $\Delta t = 380\,\text{au}$ is only ~0.08 K for OBABO and ~0.03 K for BAOAB. For comparison, that for the virial estimator is as large as ~1.3 K for OBABO and ~0.14 K for BAOAB. Fig. 5a for liquid water at $T$=300 K demonstrates that the absolute deviation for the primitive estimator at $\Delta t = 0.6\,\text{fs}$

is ~13 K for OBABO and ~0.3 K for BAOAB, while that for the virial estimator is as large as ~20

K for OBABO and ~1.3 K for BAOAB. Although the statistical error bar of the primitive estimator is about an order of magnitude larger than that of the virial estimator, the primitive estimator leads to more accurate results as the time interval $\Delta t$ of PIMD increases.

**IV. Concluding remarks**

In this paper we present a novel, simple, and accurate algorithm for implementing PIMD with the (white noise) Langevin thermostat. Its applications to the $H_2O$ molecule, liquid *para*-hydrogen and liquid water demonstrate that the BAOAB integrator for PIMD performs uniformly better than the OBABO integrator (or equivalently PILE) for configurational sampling and for calculating thermodynamic properties. Comparing to OBABO, the BAOAB integrator reduces the error by about an order of magnitude for the same time interval of PIMD, or increases the time interval by a factor of 4~6 for achieving the same convergence. Interestingly, an alternative approach to BAOAB (i.e., the BAOAB-num algorithm) can *further* improve the accuracy. (See Appendices II and III.) Although the staging transformation of path integral beads is used in PIMD for the numerical tests, the performance of BAOAB should be similar when the normal-mode transformation is employed. (See Supplemental Material[18].)

It is straightforward to extend BAOAB/BAOAB-num to imaginary time path integral based quantum dynamics methods. In comparison to OBABO, BAOAB increases the time interval for propagating the real time trajectory by a factor of 4~6 or more for the same convergence for path integral Liouville dynamics[21, 29, 33]. We expect that BAOAB/BAOAB-num also performs better than OBABO (or PILE) in thermostatted ring polymer molecular dynamics[34], centroid molecular dynamics[35], *etc*.

We note that two approaches have been proposed for combining PIMD with generalized (colored noise) Langevin thermostats[36-38]. In either approach it is demonstrated that the number of path integral beads $P$ can be decreased by a factor of 4~6 to obtain converged results for thermodynamic properties such as some structural properties and the centroid-virial version of the kinetic energy[36-38]. However, caution needs to be taken for designing estimators for other general thermodynamic properties (e.g., estimators for isotope fractionation as studied by Ceriotti and Markland[39]) in such approaches, because there is no guarantee that original estimators with color noise thermostats lead to correct results[36, 38-40]. For comparison, such as PILE or the BAOAB algorithm with the white noise Langevin dynamics for PIMD faithfully and consistently approaches exact results for any thermodynamic properties of any molecular systems as the number of beads $P$ increases. It will certainly be interesting in future work to exploit the BAOAB or BAOAB-num algorithm for PIMD with colored noise Langevin thermostats[36, 38] to achieve more efficiency, when reasonable estimators for the specific properties of interest are available.

**Acknowledgement**

This work was supported by the National Science Foundation of China (NSFC) Grants No. 21373018 and No. 21573007, by the Recruitment Program of Global Experts, by Specialized Research Fund for the Doctoral Program of Higher Education No. 20130001110009, and by Special Program for Applied Research on Super Computation of the NSFC-Guangdong Joint Fund (the second phase). We acknowledge the Beijing and Tianjin supercomputer centers for providing computational resources. This research also used resources of the National Energy Research Scientific Computing Center, a DOE Office of Science User Facility supported by the Office of Science of the U.S. Department of Energy under Contract No. DE-AC02-05CH11231.

**Appendix I: Derivation of the optimal friction coefficient**

The Langevin equation for a harmonic oscillator $V(x) = \frac{1}{2}m\omega^2 x^2$ can be expressed as

$$\ddot{x} = -\omega^2 x - \gamma_{Lang}\dot{x} + \eta(t)/m \qquad \text{(A1)}$$

with

$$\begin{aligned} \langle \eta(t) \rangle &= 0 \\ \langle \eta(t)\eta(t') \rangle &= \frac{2\gamma_{Lang}m}{\beta}\delta(t-t') \end{aligned} \qquad \text{(A2)}$$

as time averages over an infinitesimal time interval.

Implementing the Laplace transform $\hat{f}(z) = L[f(t)] = \int_0^\infty e^{-zt} f(t)\,dt$ to Eq. (A1) leads to

$$\hat{x}(z) = \frac{zx(0)}{z^2 + \gamma_{Lang}z + \omega^2} + \frac{\gamma_{Lang}x(0) + \dot{x}(0)}{z^2 + \gamma_{Lang}z + \omega^2} + \frac{\hat{\eta}(z)/m}{z^2 + \gamma_{Lang}z + \omega^2} \quad . \qquad \text{(A3)}$$

The inverse Laplace transform of Eq. (A3) then produces

$$x(t) = x(0)\frac{z_1 e^{z_1 t} - z_2 e^{z_2 t}}{z_1 - z_2} + \left[ \gamma_{Lang}x(0) + \frac{p(0)}{m} \right]\frac{e^{z_1 t} - e^{z_2 t}}{z_1 - z_2} + \frac{1}{m}\int_0^t \eta(s)\frac{e^{z_1(t-s)} - e^{z_2(t-s)}}{z_1 - z_2}\,ds \qquad \text{(A4)}$$

with

$$\begin{aligned} z_1 &= -\frac{\gamma_{Lang}}{2} + \frac{1}{2}\sqrt{\gamma_{Lang}^2 - 4\omega^2} \\ z_2 &= -\frac{\gamma_{Lang}}{2} - \frac{1}{2}\sqrt{\gamma_{Lang}^2 - 4\omega^2} \end{aligned} \qquad . \qquad \text{(A5)}$$

Because the normalized phase space probability distribution $\rho(x, p)$ generated from the Langevin equation Eq. (A1) is

$$\frac{\beta\omega}{2\pi}\exp\left[-\beta\left(\frac{p^2}{2m}+\frac{1}{2}m\omega^2 x^2\right)\right] \quad,$$

(A6)

it is straightforward to verify

$$
\begin{aligned}
m^2\beta^2\omega^4\left\langle x^2(0)x^2(t)\right\rangle = {} & 1 - \frac{\omega^2}{\lambda^2}e^{-\gamma_{Lang}t} \\
& + \frac{\omega^2+2\lambda^2+\lambda\gamma_{Lang}}{2\lambda^2}e^{(2\lambda-\gamma_{Lang})t} \\
& + \frac{\omega^2+2\lambda^2-\lambda\gamma_{Lang}}{2\lambda^2}e^{(-2\lambda-\gamma_{Lang})t}
\end{aligned}
$$

(A7)

with $\lambda=(z_1-z_2)/2$ . The characteristic correlation time of the potential energy autocorrelation function

$$C_{pot}(t)=\frac{\left\langle\left[V(x(t))-\langle V(x)\rangle\right]\left[V(x(0))-\langle V(x)\rangle\right]\right\rangle}{\left\langle\left[V(x)-\langle V(x)\rangle\right]^2\right\rangle}$$

(A8)

can then be shown as

$$\tau_{pot}=\int_0^\infty C_{pot}(t)\,dt=\frac{1}{2}\left(\frac{1}{\gamma_{Lang}}+\frac{\gamma_{Lang}}{\omega^2}\right) \quad.$$

(A9)

The smaller $\tau_{pot}$ is, the more efficiently the Langevin equation explores the potential energy surface and samples the configuration space. When $\gamma_{Lang}=\omega$ , the characteristic correlation time $\tau_{pot}$ reaches the minimum value

$$\tau_{pot}^{\min}=\frac{1}{\omega} \quad.$$

(A10)

Since $\phi = 0$ in Eq. (25) in the free particle limit, the equations of motion for the staging modes $\left( j = \overline{2,P} \right)$ are decoupled. The Langevin equation of each degree of freedom $\left( i = \overline{1,N} \right)$ of each staging mode $\left( j = \overline{2,P} \right)$ in Eq. (25) then shares the same form as Eq. (A1). This leads to Eq. (26) for the optimal Langevin friction coefficients for PIMD with the staging transformation, because only the configurational distribution of PIMD is useful.

Ceriotti *et al.* were the first to exploit an analytical knowledge of the path integral normal mode frequencies in the free particle limit for choosing the optimal friction coefficients[17]. Here the similar strategy is employed for staging PIMD.

**Appendix II: A similar approach to BAOAB or OBABO**

Langevin dynamics Eq. (25) can also be divided into three parts in an alternative way

$$\begin{pmatrix} \dot{\boldsymbol{\xi}} \\ \dot{\mathbf{p}} \end{pmatrix} = \underbrace{\begin{pmatrix} \tilde{\mathbf{M}}^{-1}\,\mathbf{p} \\ 0 \end{pmatrix}}_{A} + \underbrace{\begin{pmatrix} 0 \\ -\omega_P^2\overline{\mathbf{M}}\boldsymbol{\xi} - \dfrac{\partial\phi}{\partial\boldsymbol{\xi}} \end{pmatrix}}_{B} + \underbrace{\begin{pmatrix} 0 \\ -\boldsymbol{\gamma}_{Lang}\mathbf{p} + \boldsymbol{\sigma}\,\tilde{\mathbf{M}}^{1/2}\boldsymbol{\eta}(t) \end{pmatrix}}_{O} \; . \tag{A11}$$

That is, the force term $-\omega_P^2\overline{\mathbf{M}}\boldsymbol{\xi}$ is moved from part A to part B in Eq. (45). The analytic propagator of the harmonic part of Eq. (45) is no longer employed.

When the splitting Eq. (38) is used for Eq. (A11), we note the new integrator BAOAB-num. Similar to Section II-3, such an algorithm for propagating the PIMD trajectory through a time interval $\Delta t$ for Eq. (A11) is

$$\mathbf{p}_j \leftarrow \mathbf{p}_j - \frac{\partial\phi}{\partial\boldsymbol{\xi}_j}\frac{\Delta t}{2} - \omega_P^2\overline{\mathbf{M}}_j\boldsymbol{\xi}_j\,\frac{\Delta t}{2} \qquad \left( j = \overline{1,P} \right) \tag{A12}$$

$$\xi_j \leftarrow \xi_j + \tilde{\mathbf{M}}_j^{-1}\mathbf{p}_j\frac{\Delta t}{2} \qquad \left(j = \overline{1,P}\right) \tag{A13}$$

$$\mathbf{p}_j \leftarrow \tilde{c}_1^{(j)}\mathbf{p}_j + \tilde{c}_2^{(j)}\sqrt{\frac{1}{\beta}}\left(\tilde{\mathbf{M}}_j\right)^{1/2}\boldsymbol{\eta}_j \qquad \left(j = \overline{1,P}\right) \tag{A14}$$

$$\xi_j \leftarrow \xi_j + \tilde{\mathbf{M}}_j^{-1}\mathbf{p}_j\frac{\Delta t}{2} \qquad \left(j = \overline{1,P}\right) \tag{A15}$$

$$\mathbf{p}_j \leftarrow \mathbf{p}_j - \frac{\partial \phi}{\partial \xi_j}\frac{\Delta t}{2} - \omega_P^2\bar{\mathbf{M}}_j\xi_j\frac{\Delta t}{2} \qquad \left(j = \overline{1,P}\right) \quad . \tag{A16}$$

Here $\boldsymbol{\eta}_j$, $\tilde{c}_1^{(j)}$ and $\tilde{c}_2^{(j)}$ are defined in the same way as in BAOAB [Eqs. (39)-(43)]. Following the same procedure as shown in Eqs. (53)-(70) of Section II-4, it is straightforward to verify that the configurational distribution produced by BAOAB-num is

$$\rho_{BAOAB-num}^{config}\left(\xi\right) = \rho_{eq}^{config}\left(\xi\right)\left\{1 + \beta\omega_P^2\Delta t^2\left[O(\varepsilon)\right] + O(\omega_P^4\Delta t^4)\right\} \quad , \tag{A17}$$

while truncating at the 0-th order of $\varepsilon$ for the term associated with $\omega_P^2\Delta t^2$. It is further shown in Appendix III that BAOAB-num leads to the exact configurational distribution of the beads in the harmonic limit, regardless of the time interval $\Delta t$ and the friction coefficients $\gamma_{Lang}$.

Similarly, when the splitting Eq. (31) is employed for Eq. (A11), we note the new integrator OBABO-num. Such an algorithm for propagating the PIMD trajectory through a time interval $\Delta t$ for Eq. (A11) is

$$\mathbf{p}_j \leftarrow c_1^{(j)}\mathbf{p}_j + c_2^{(j)}\sqrt{\frac{1}{\beta}}\left(\tilde{\mathbf{M}}_j\right)^{1/2}\boldsymbol{\eta}_j \qquad \left(j = \overline{1,P}\right) \tag{A18}$$

$$\mathbf{p}_j \leftarrow \mathbf{p}_j - \frac{\partial \phi}{\partial \boldsymbol{\xi}_j}\frac{\Delta t}{2} - \omega_P^2 \bar{\mathbf{M}}_j \boldsymbol{\xi}_j \frac{\Delta t}{2} \qquad \left(j = \overline{1,P}\right) \tag{A19}$$

$$\boldsymbol{\xi}_j \leftarrow \boldsymbol{\xi}_j + \tilde{\mathbf{M}}_j^{-1}\mathbf{p}_j \Delta t \qquad \left(j = \overline{1,P}\right) \tag{A20}$$

$$\mathbf{p}_j \leftarrow \mathbf{p}_j - \frac{\partial \phi}{\partial \boldsymbol{\xi}_j}\frac{\Delta t}{2} - \omega_P^2 \bar{\mathbf{M}}_j \boldsymbol{\xi}_j \frac{\Delta t}{2} \qquad \left(j = \overline{1,P}\right) \tag{A21}$$

$$\mathbf{p}_j \leftarrow c_1^{(j)}\mathbf{p}_j + c_2^{(j)}\sqrt{\frac{1}{\beta}}\left(\tilde{\mathbf{M}}_j\right)^{1/2}\boldsymbol{\eta}_j \qquad \left(j = \overline{1,P}\right) \quad . \tag{A22}$$

Here $\boldsymbol{\eta}_j$, $c_1^{(j)}$ and $c_2^{(j)}$ are defined in the same way as in OBABO [Eqs. (32)-(36)]. Following the same procedure as shown in Eqs. (53)-(70) of Section II-4, one finds

$$\rho_{OBABO-num}^{config}(\boldsymbol{\xi}) = \rho_{eq}^{config}(\boldsymbol{\xi})\left\{1 + \beta\omega_P^2\Delta t^2\left[\frac{1}{\varepsilon^2}\left(\frac{1}{8}\boldsymbol{\xi}^T\bar{\mathbf{M}}\tilde{\mathbf{M}}^{-1}\bar{\mathbf{M}}\boldsymbol{\xi} + \bar{\bar{G}}_{-2}\right)\right.\right.$$
$$\left.\left. + \left(\frac{1}{4}\boldsymbol{\xi}^T\bar{\mathbf{M}}\tilde{\mathbf{M}}^{-1}\frac{\partial \phi}{\partial \boldsymbol{\xi}} + \bar{\bar{G}}_0\right) + O(\varepsilon)\right] + O(\omega_P^4\Delta t^4)\right\}, \tag{A23}$$

while truncating at the 0-th order of $\varepsilon$ for the term associated with $\omega_P^2\Delta t^2$. Here $\bar{\bar{G}}_{-2}$ and $\bar{\bar{G}}_0$ are two constants required for normalization. It is easy to show that Eq. (A23) for the harmonic system [Eq. (71) or Eq. (72)] becomes

$$
\rho_{OBABO-num}^{config}\left(\xi\right) = \rho_{eq}^{config}\left(\xi\right)
$$

$$
\times \left\{ 1 + \beta \omega_p^2 \Delta t^2 \left[ \frac{1}{\varepsilon^2} \left( \frac{1}{8} \xi^T \bar{\mathbf{M}} \tilde{\mathbf{M}}^{-1} \bar{\mathbf{M}} \xi - \frac{1}{\beta} \frac{1}{8} \mathrm{Tr} \left( \bar{\mathbf{M}} \tilde{\mathbf{M}}^{-1} \bar{\mathbf{M}} \left( \omega_p^2 \bar{\mathbf{M}} + \mathbf{K} \right)^{-1} \right) \right) \right. \right.
$$

$$
\left. + \left( \frac{1}{4} \xi^T \bar{\mathbf{M}} \tilde{\mathbf{M}}^{-1} \mathbf{K} (\xi - \xi_{eq}) - \frac{1}{\beta} \frac{1}{4} \mathrm{Tr} \left( \bar{\mathbf{M}} \tilde{\mathbf{M}}^{-1} \mathbf{K} \left( \omega_p^2 \bar{\mathbf{M}} + \mathbf{K} \right)^{-1} \right) \right) + O(\varepsilon) \right]
$$

$$
\left. + O(\omega_p^4 \Delta t^4) \right\} \qquad \text{.(A24)}
$$

Appendix III further demonstrates that the analytical form of the steady state of OBABO-num in the harmonic limit can actually be obtained. It also shows that OBABO-num does *not* even produce the exact configurational distribution for the beads in the free particle limit when the time interval $\Delta t$ is finite, while the other three integrators (BAOAB, BAOAB-num, and OBABO) are able to do so. Below we compare the integrators using two examples.

Consider the 1-dimensional model potential $V(x) = 0.25x^4$. Since the potential has no harmonic term, it presents a good test to compare BAOAB-num to the other integrators. Use $P = 1024$ beads for the inverse temperature $\beta = 8$. Fig. 6 depicts the results for the average potential energy using different time intervals of PIMD, where the four integrators (BAOAB, OBABO, BAOAB-num, and OBABO-num) are compared. In agreement with the analysis in Section II-4 and that in Appendix III, Fig. 6 shows that BAOAB-num is the most accurate algorithm while OBABO-num is the least. Both BAOAB and BAOAB-num demonstrates a significantly better performance than OBABO and OBABO-num.

We further compare BAOAB and BAOAB-num using the $H_2O$ molecule at $T = 300\,\mathrm{K}$. The parameters are the same as those listed in Section III-1. Fig. 7 compares the BAOAB results and the BAOAB-num ones for the average potential energy per atom using different time intervals of PIMD. When $\Delta t$ is smaller than 18 au, BAOAB-num is more accurate than BAOAB. The two integrators produce almost the same results for $\Delta t$ is close to 18 au. BAOAB-num is less stable than BAOAB as the time interval $\Delta t$ increases in the region $\Delta t > 18$ au, while both BAOAB-num and BAOAB fail when $\Delta t$ is greater than 20 au, (as shown in Fig. 7b where the horizontal axis represents $\log \left| \langle V(\hat{\mathbf{x}}) \rangle \big/ \left( N_{atom} k_B \right) \right|$). When $P$ or $\omega_P$ is significantly large, it is expected that BAOAB-num is less stable as the time interval $\Delta t$ is considerably large. This is because the dominate harmonic force term is analytically integrated in part A of Eq. (45), while the same term is numerically solved in part B of Eq. (A11). It is worth emphasizing that results are not converged at all in the region where BAOAB is more accurate than BAOAB-num. Fig. 6 does not even show such a region before both integrators fail. Figs. 6 and 7 suggest that BAOAB-num always demonstrates better performance in the region where BAOAB produces reasonably converged results.

Comparison of the results for the average kinetic energy also demonstrates the same trend, which is not shown here.

## Appendix III: Accuracy of the PIMD integrator in the harmonic limit and in the free particle limit

As discussed in Section II-4, while both BAOAB and OBABO are exact in the free particle limit, neither of them produces the exact configurational distribution for the path integral beads [Eq. (69)] in the harmonic limit when the time interval $\Delta t$ is finite.

Below we investigate the accuracy of BAOAB-num and that of OBABO-num proposed in Appendix II for the (general) harmonic system [Eq. (71)].

When the splitting Eq. (38) is used for Eq. (A11), the phase space propagator for BAOAB-num in the time interval $\Delta t$ is

$$e^{\mathscr{L}^{BAOAB-num}\Delta t} = e^{\mathscr{L}_B^{num}\Delta t/2}e^{\mathscr{L}_A^{num}\Delta t/2}e^{\mathscr{L}_O^{num}\Delta t}e^{\mathscr{L}_A^{num}\Delta t/2}e^{\mathscr{L}_B^{num}\Delta t/2} \qquad , \tag{A25}$$

where

$$\mathscr{L}_A^{num}\rho = -\mathbf{p}^T\tilde{\mathbf{M}}^{-1}\frac{\partial \rho}{\partial \boldsymbol{\xi}} \qquad , \tag{A26}$$

$$\mathscr{L}_B^{num}\rho = \omega_P^2\boldsymbol{\xi}^T\bar{\mathbf{M}}\frac{\partial \rho}{\partial \mathbf{p}} + \left(\frac{\partial \phi}{\partial \boldsymbol{\xi}}\right)^T\frac{\partial \rho}{\partial \mathbf{p}} = \left(\frac{\partial U_{eff}}{\partial \boldsymbol{\xi}}\right)^T\frac{\partial \rho}{\partial \mathbf{p}} \qquad , \tag{A27}$$

$$\mathscr{L}_O^{num}\rho = \frac{\partial}{\partial \mathbf{p}}\cdot\left(\boldsymbol{\gamma}_{Lang}\mathbf{p}\rho\right) + \frac{1}{\beta}\frac{\partial}{\partial \mathbf{p}}\cdot\left(\boldsymbol{\gamma}_{Lang}\tilde{\mathbf{M}}\frac{\partial \rho}{\partial \mathbf{p}}\right) \qquad . \tag{A28}$$

Consider the harmonic system Eq. (71), which leads to Eq. (78) and Eq. (79). Eq. (A27) then becomes

$$\mathscr{L}_B^{num}\rho = (\boldsymbol{\xi}-\boldsymbol{\xi}_{eq})^T\boldsymbol{\Omega}\frac{\partial \rho}{\partial \mathbf{p}} \qquad . \tag{A29}$$

It is straightforward to show that the Taylor expansion $e^{\mathscr{L}_A^{num}\Delta t/2} = \sum_{n=0}^{\infty}\frac{1}{n!}\left(-\mathbf{p}^T\tilde{\mathbf{M}}^{-1}\frac{\Delta t}{2}\frac{\partial}{\partial \boldsymbol{\xi}}\right)^n$

leads to

$$e^{\mathscr{L}_A^{num}\Delta t/2}f(\boldsymbol{\xi}) = f(\boldsymbol{\xi}-\tilde{\mathbf{M}}^{-1}\mathbf{p}\frac{\Delta t}{2}) \qquad . \tag{A30}$$

Similarly, one obtains

$$e^{\mathscr{L}_B^{num}\Delta t/2}g(\mathbf{p}) = g\left(\mathbf{p}+\boldsymbol{\Omega}(\boldsymbol{\xi}-\boldsymbol{\xi}_{eq})\frac{\Delta t}{2}\right) \qquad . \tag{A31}$$

The OU process keeps the Maxwell momentum distribution unchanged, i.e.,

$$e^{\mathcal{L}_O^{num}\Delta t}\exp\left\{-\beta\left[\frac{1}{2}\mathbf{p}^T\tilde{\mathbf{M}}^{-1}\mathbf{p}\right]\right\}=\exp\left\{-\beta\left[\frac{1}{2}\mathbf{p}^T\tilde{\mathbf{M}}^{-1}\mathbf{p}\right]\right\}\quad. \tag{A32}$$

Consider the density distribution

$$\rho^{BAOAB-num}\left(\boldsymbol{\xi},\mathbf{p}\right)=\frac{1}{Z_N}\exp\left\{-\beta\left[\frac{1}{2}\mathbf{p}^T\left(\tilde{\mathbf{M}}-\boldsymbol{\Omega}\frac{\Delta t^2}{4}\right)^{-1}\mathbf{p}+\frac{1}{2}(\boldsymbol{\xi}-\boldsymbol{\xi}_{eq})^T\boldsymbol{\Omega}(\boldsymbol{\xi}-\boldsymbol{\xi}_{eq})\right]\right\}\quad, \tag{A33}$$

where $Z_N$ is the normalization coefficient. Using Eqs. (A25) and (A30)-(A32), it is easy to verify

$$e^{\mathcal{L}^{BAOAB-num}\Delta t}\rho^{BAOAB-num}=\rho^{BAOAB-num}\quad. \tag{A34}$$

That is, Eq. (A33) is a steady state of the BAOAB-num integrator. Integration over $\mathbf{p}$ in Eq. (A33) leads to

$$\begin{aligned}\rho_{BAOAB-num}^{config}\left(\boldsymbol{\xi}\right)&=\frac{1}{Z_N'}\exp\left\{-\beta\left(\frac{1}{2}(\boldsymbol{\xi}-\boldsymbol{\xi}_{eq})^T\boldsymbol{\Omega}(\boldsymbol{\xi}-\boldsymbol{\xi}_{eq})\right)\right\}\\&=\frac{1}{Z_N'}\exp\left[-\beta U_{eff}\left(\boldsymbol{\xi}\right)\right]\\&=\rho_{eq}^{config}\left(\boldsymbol{\xi}\right)\end{aligned}\quad, \tag{A35}$$

where $Z_N'$ is the new normalization coefficient. The BAOAB-num integrator in principle leads to the exact configurational distribution [Eq. (69)] (of the path integral beads) for the harmonic system (which includes the free particle case), irrespective of the time interval $\Delta t$ (as long as the propagation Eq. (A25) is numerically stable) and the friction coefficients $\gamma_{Lang}$.

When the number of path integral beads $P\to 1$, i.e., PIMD reduces to classical MD, BAOAB and BAOAB-num are the same in the classical limit. Eq. (A35) then suggests that the BAOAB/BAOAB-num thermostatting algorithm for classical MD leads to the *exact* classical configurational distribution for the harmonic system, regardless of the time interval $\Delta t$ and the Langevin friction coefficient. Because the proof [Eqs. (A25)-(A35)] involves no approximation, this remarkable conclusion complements the analysis on BAOAB for classical MD in the large friction limit by Leimkuhler and Matthews[26-28].

Similarly, the steady density distribution for the OBABO-num integrator for the harmonic system is

$$\rho^{OBABO-num} = \frac{1}{\bar{Z}_N} \exp\left[ -\beta \left( \frac{1}{2} \mathbf{p}^T \tilde{\mathbf{M}}^{-1} \mathbf{p} + \frac{1}{2} (\xi - \xi_{eq})^T (\mathbf{1} - \mathbf{\Omega} \tilde{\mathbf{M}}^{-1} \frac{\Delta t^2}{4}) \mathbf{\Omega} (\xi - \xi_{eq}) \right) \right] \quad , \quad (A36)$$

which produces the configurational distribution

$$\rho^{config}_{OBABO-num} = \frac{1}{\bar{Z}_N'} \exp\left\{ -\beta \left[ \frac{1}{2} (\xi - \xi_{eq})^T \left( \mathbf{1} - \mathbf{\Omega} \tilde{\mathbf{M}}^{-1} \frac{\Delta t^2}{4} \right) \mathbf{\Omega} (\xi - \xi_{eq}) \right] \right\} \quad . \quad (A37)$$

Here $\bar{Z}_N$ and $\bar{Z}_N'$ are the normalization coefficients. One expands the density into a power series of $\Delta t$ or $\omega_p \Delta t$ and then finds

$$
\begin{aligned}
\rho^{config}_{OBABO-num} (\xi) = {} & \rho^{config}_{eq} (\xi) \\
& \times \Bigg\{ 1 + \beta \omega_p^2 \Delta t^2 \Bigg[ \omega_p^2 \left( \frac{1}{8} \xi^T \bar{\mathbf{M}} \tilde{\mathbf{M}}^{-1} \bar{\mathbf{M}} \xi - \frac{1}{\beta} \frac{1}{8} \mathrm{Tr} \left( \bar{\mathbf{M}} \tilde{\mathbf{M}}^{-1} \bar{\mathbf{M}} \left( \omega_p^2 \bar{\mathbf{M}} + \mathbf{K} \right)^{-1} \right) \right) \\
& + \left( \frac{1}{4} \xi^T \bar{\mathbf{M}} \tilde{\mathbf{M}}^{-1} \mathbf{K} (\xi - \xi_{eq}) - \frac{1}{\beta} \frac{1}{4} \mathrm{Tr} \left( \bar{\mathbf{M}} \tilde{\mathbf{M}}^{-1} \mathbf{K} \left( \omega_p^2 \bar{\mathbf{M}} + \mathbf{K} \right)^{-1} \right) \right) \\
& + \varepsilon^2 \left( \frac{1}{8} (\xi - \xi_{eq})^T \mathbf{K} \tilde{\mathbf{M}}^{-1} \mathbf{K} (\xi - \xi_{eq}) - \frac{1}{\beta} \frac{1}{8} \mathrm{Tr} \left( \mathbf{K} \tilde{\mathbf{M}}^{-1} \mathbf{K} \left( \omega_p^2 \bar{\mathbf{M}} + \mathbf{K} \right)^{-1} \right) \right) \Bigg] \\
& + O(\omega_p^4 \Delta t^4) \Bigg\}
\end{aligned}
\quad . (A38)
$$

While truncating at the 0-th order of $\varepsilon$ for the term associated with $\omega_p^2 \Delta t^2$, Eq. (A38) becomes

$$
\begin{aligned}
\rho_{OBABO-num}^{config}\left(\xi\right) = {} & \rho_{eq}^{config}\left(\xi\right) \\
& \times \left\{ 1 + \beta\omega_P^2\Delta t^2 \left[ \frac{1}{\varepsilon^2}\left( \frac{1}{8}\xi^T\bar{\mathbf{M}}\tilde{\mathbf{M}}^{-1}\bar{\mathbf{M}}\xi - \frac{1}{\beta}\frac{1}{8}\mathrm{Tr}\left( \bar{\mathbf{M}}\tilde{\mathbf{M}}^{-1}\bar{\mathbf{M}}\left( \omega_P^2\bar{\mathbf{M}} + \mathbf{K} \right)^{-1} \right) \right) \right. \right. \\
& \left. \left. + \left( \frac{1}{4}\xi^T\bar{\mathbf{M}}\tilde{\mathbf{M}}^{-1}\mathbf{K}(\xi - \xi_{eq}) - \frac{1}{\beta}\frac{1}{4}\mathrm{Tr}\left( \bar{\mathbf{M}}\tilde{\mathbf{M}}^{-1}\mathbf{K}\left( \omega_P^2\bar{\mathbf{M}} + \mathbf{K} \right)^{-1} \right) \right) + O(\varepsilon^2) \right] \right. \\
& \left. + O(\omega_P^4\Delta t^4) \right\}
\end{aligned} \qquad \text{(A39)}
$$

It is consistent with Eq. (A24) except that Eq. (A39) states that the accuracy is now up to $O(\varepsilon^2)$ for the term associated with $\omega_P^2\Delta t^2$. The governing term of the error of the configurational distribution is then $\frac{\beta}{8}\xi^T\bar{\mathbf{M}}\tilde{\mathbf{M}}^{-1}\bar{\mathbf{M}}\xi\omega_P^4\Delta t^2$. Because $\omega_P$ is often large for converged PIMD results, comparing Eq. (A39) (or Eq. (A24)) to Eqs. (80), (82) and (A35), one finds the ascending order for the error of the configurational distribution (in the harmonic limit):

$$
\text{BAOAB-num} < \text{BAOAB} < \text{OBABO} < \text{OBABO-num} \qquad . \qquad \text{(A40)}
$$

Finally, we consider the free particle system where $\phi = 0$. It is trivial to obtain

$$
\mathbf{\Omega} = \omega_P^2\bar{\mathbf{M}} \qquad \text{(A41)}
$$

from Eq. (79). Inserting Eq. (A41) and Eq. (77) into Eq. (A37), one obtains

$$
\begin{aligned}
\rho_{OBABO-num}^{config} = {} & \frac{1}{Vol_N}\left( \frac{\beta\omega_P^2}{2\pi} \right)^{N(P-1)/2}\left| \prod_{j=2}^{P}\det\left( \bar{\mathbf{M}}_j \right) \right|^{1/2}\left| \det\left( \mathbf{1} - \omega_P^2\bar{\mathbf{M}}\tilde{\mathbf{M}}^{-1}\frac{\Delta t^2}{4} \right) \right|^{1/2} \times \\
& \exp\left\{ -\beta\left[ \frac{\omega_P^2}{2}\xi^T\left( \mathbf{1} - \omega_P^2\bar{\mathbf{M}}\tilde{\mathbf{M}}^{-1}\frac{\Delta t^2}{4} \right)\bar{\mathbf{M}}\xi \right] \right\}
\end{aligned} \qquad \text{(A42)}
$$

Here $Vol_N = \int d\xi_1 = \int d\mathbf{x}_1$ represents the volume of the system. Expanding the density into a power series of $\Delta t$ or $\omega_P\Delta t$, one finds

$$\begin{aligned}
\rho_{OBABO-num}^{config} = \frac{1}{Vol_N} & \left( \frac{\beta \omega_P^2}{2\pi} \right)^{N(P-1)/2} \left| \prod_{j=2}^{P} \det \left( \bar{\mathbf{M}}_j \right) \right|^{1/2} \exp \left\{ -\beta \left[ \frac{\omega_P^2}{2} \boldsymbol{\xi}^T \bar{\mathbf{M}} \boldsymbol{\xi} \right] \right\} \\
& \times \left( 1 + \beta \omega_P^2 \Delta t^2 \left[ \frac{1}{8} \omega_P^2 \boldsymbol{\xi}^T \bar{\mathbf{M}} \tilde{\mathbf{M}}^{-1} \bar{\mathbf{M}} \boldsymbol{\xi} - \frac{1}{8\beta} \mathrm{Tr} \left( \bar{\mathbf{M}} \tilde{\mathbf{M}}^{-1} \right) \right] + O \left( \omega_P^4 \Delta t^4 \right) \right)
\end{aligned} \qquad . \quad \text{(A43)}$$

Eqs. (A37) and (A43) suggest that OBABO-num does *not* lead to the exact configurational distribution of the path integral beads even in the free particle limit, while all other three integrators (BAOAB, BAOAB-num, and OBABO) do so.

Conclusions are similar when the integrators are extended to normal mode PIMD. (See Supplemental Material[18].)

**Figure Captions**

**Fig. 1** (Color). PIMD results using different time intervals for $H_2O$ at T=300K. (a) The average kinetic energy per atom $\langle \hat{\mathbf{p}}^T \mathbf{M}^{-1} \hat{\mathbf{p}} \rangle / (2 N_{atom} k_B)$ (unit: Kelvin). Both primitive and virial estimators are used. (b) Difference between the primitive and virial estimators. (unit: Kelvin). (c) The average potential energy per atom $\langle V(\hat{\mathbf{x}}) \rangle / (N_{atom} k_B)$ (unit: Kelvin). Solid line: BAOAB results. Dotted line: OBABO results. The unit of the time interval is atomic unit (au). Statistical error bars are included.

**Fig. 2** (Color). As in Fig. 1, but for $H_2O$ at T=100K.

**Fig. 3** (Color). PIMD results using different time intervals for liquid *para*-hydrogen at T=25K. (a) The average kinetic energy per molecule $\langle \hat{\mathbf{p}}^T \mathbf{M}^{-1} \hat{\mathbf{p}} \rangle / (2 N_{mol} k_B)$ (unit: Kelvin). Both primitive and virial estimators are used. (b) Difference between the primitive and virial estimators. (unit: Kelvin). (c) The average potential energy per molecule $\langle V(\hat{\mathbf{x}}) \rangle / (N_{mol} k_B)$ (unit: Kelvin). Solid line: BAOAB results. Dotted line: OBABO results. The unit of the time interval is atomic unit (au). Statistical error bars are included.

**Fig. 4** (Color). As in Fig. 3, but for liquid *para*-hydrogen at T=14K.

**Fig. 5** (Color). PIMD results using different time intervals for liquid water at T=300K. (a) The average kinetic energy per atom $\langle \hat{\mathbf{p}}^T \mathbf{M}^{-1} \hat{\mathbf{p}} \rangle / (2 N_{atom} k_B)$ (unit: Kelvin). Both primitive and virial estimators are used. (b) Difference between the primitive and virial estimators. (unit: Kelvin). (c) The average potential energy per atom $\langle V(\hat{\mathbf{x}}) \rangle / (N_{atom} k_B)$ (unit: Kelvin). Solid line: BAOAB results. Dotted line: OBABO results. The unit of the time interval is femtosecond (fs). Statistical error bars are included.

**Fig. 6** (Color). PIMD results for the average potential energy $\langle V(\hat{\mathbf{x}}) \rangle$ using different time intervals for $V(x) = 0.25x^4$ at $\beta = 8$. Solid line: BAOAB and BAOAB-num results. Dotted line: OBABO and OBABO-num results. The unit of either the potential energy or the time interval is atomic unit (au). Statistical error bars are included.

**Fig. 7** (Color). PIMD results using different time intervals for $H_2O$ at T=300K. (a) The average potential energy per atom $\langle V(\hat{\mathbf{x}}) \rangle / (N_{atom} k_B)$ (unit: Kelvin). (b) Logarithm of the average potential energy per atom $\log \left| \langle V(\hat{\mathbf{x}}) \rangle / (N_{atom} k_B) \right|$. Solid line: BAOAB results. Dashed line: BAOAB-num results. The unit of the time interval is atomic unit (au). Statistical error bars are included.

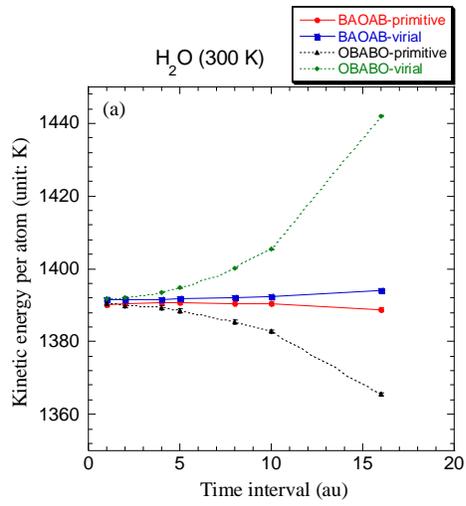

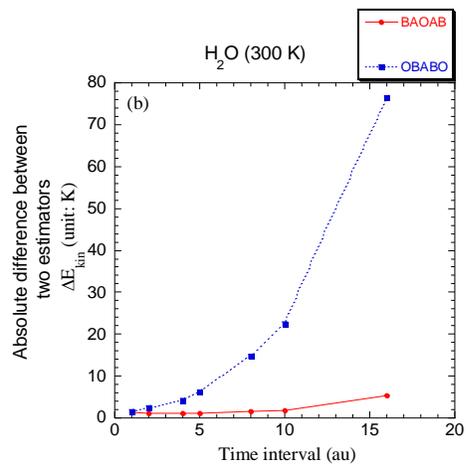

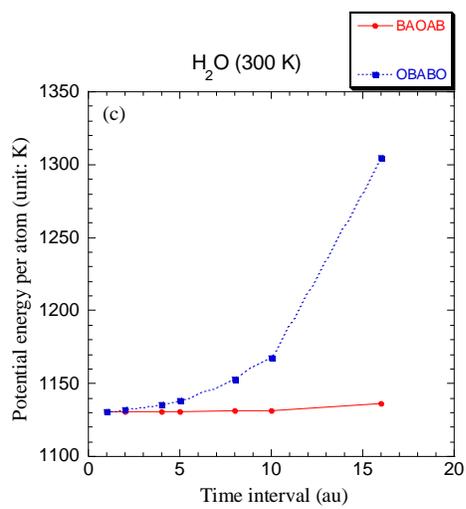

**Fig. 1**

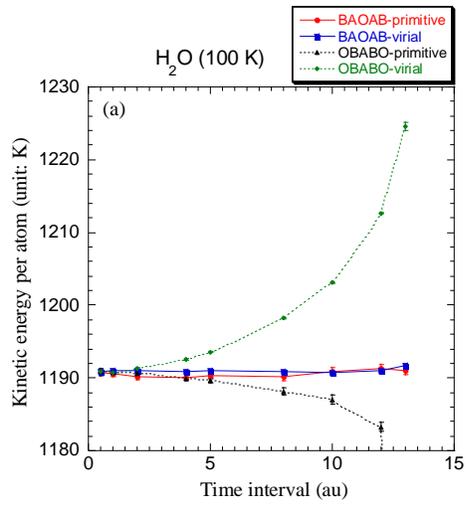

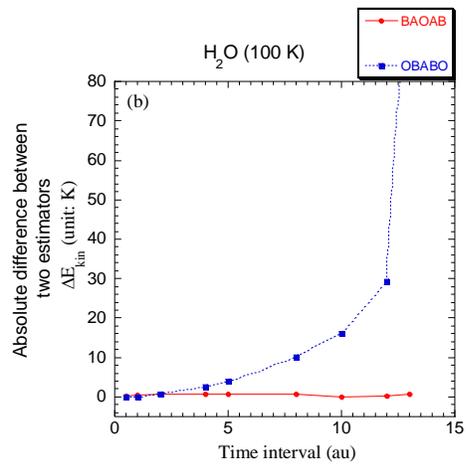

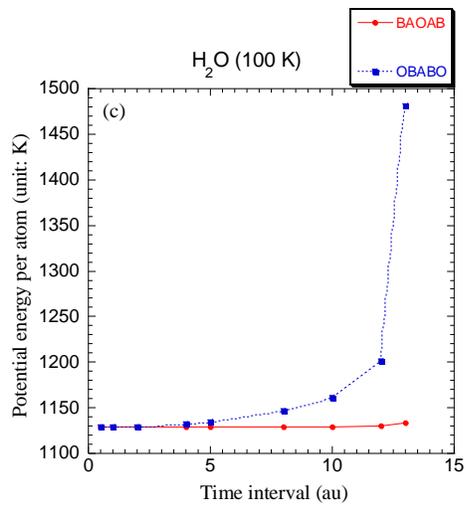

**Fig. 2**

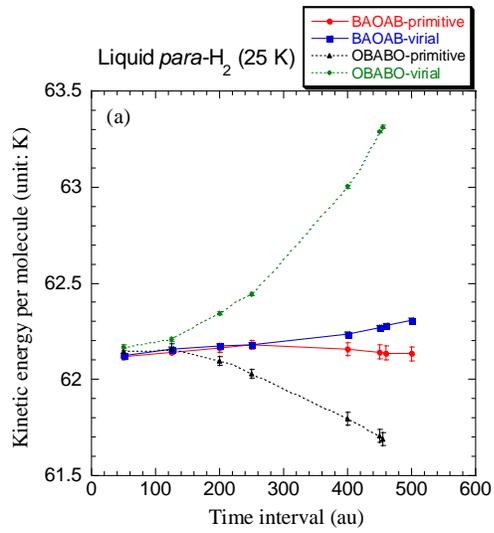

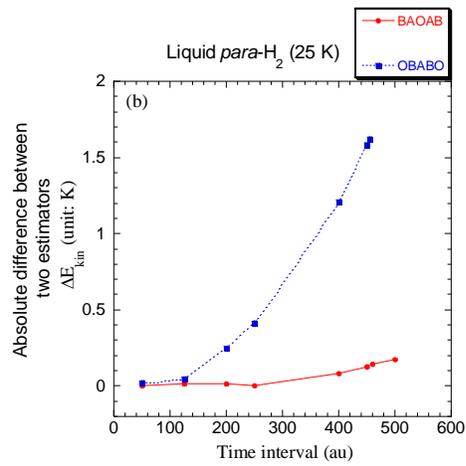

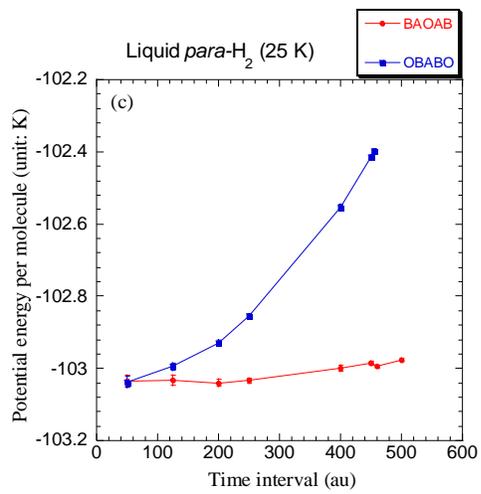

**Fig. 3**

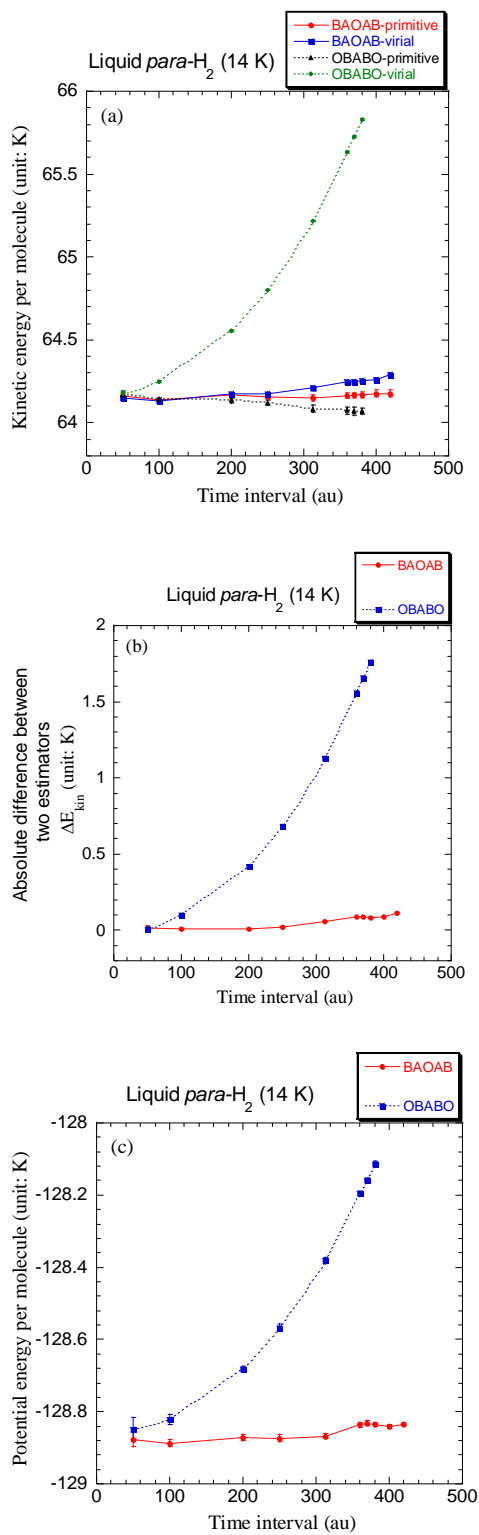

Liquid *para*-H$_2$ (14 K)

(a)

BAOAB-primitive
BAOAB-virial
OBABO-primitive
OBABO-virial

Liquid *para*-H$_2$ (14 K)

(b)

BAOAB
OBABO

Liquid *para*-H$_2$ (14 K)

(c)

BAOAB
OBABO

**Fig. 4**

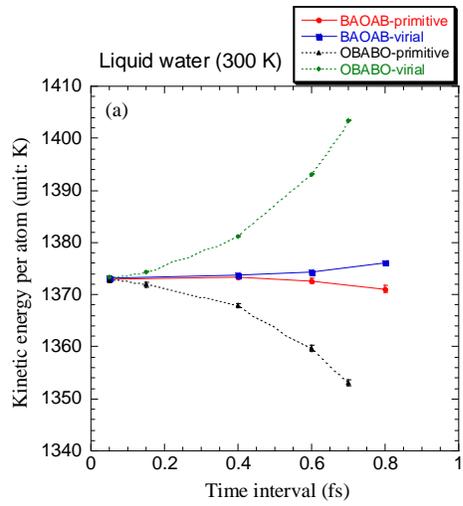

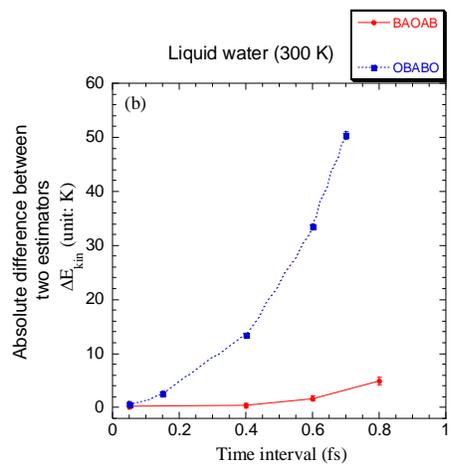

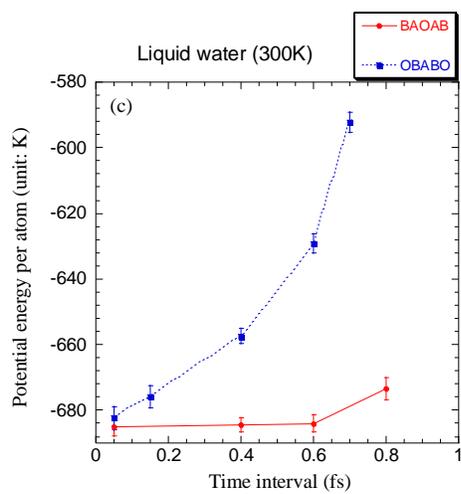

**Fig. 5**

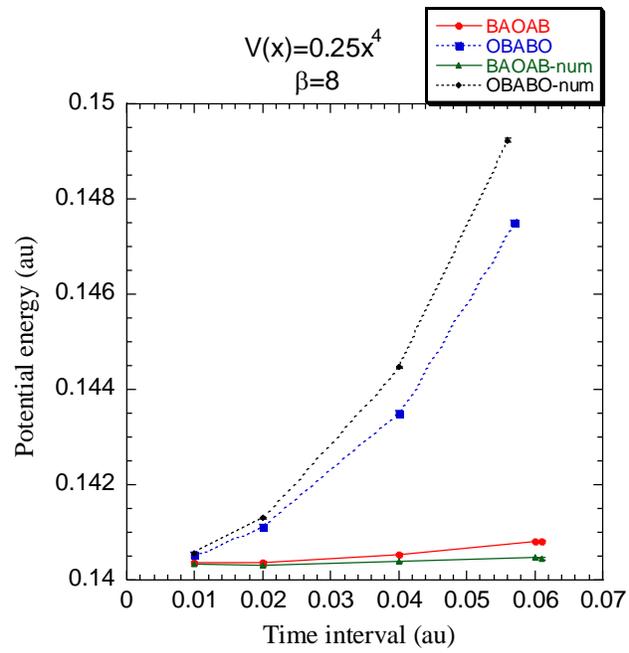

**Fig. 6**

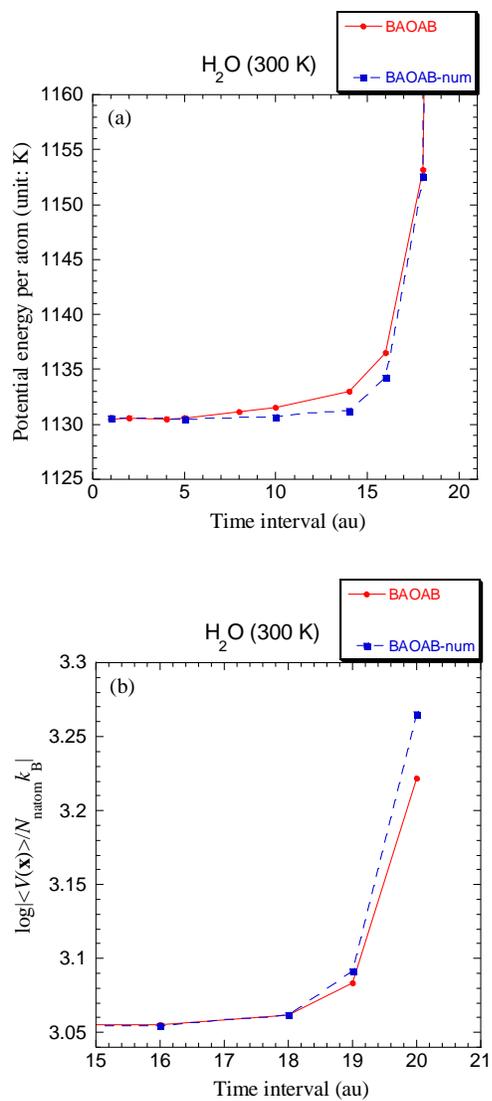

**Fig. 7**

**Supplemental material for**

**A simple and accurate integrator for path integral molecular dynamics**

**with the Langevin thermostat**

Jian Liu [1, a)], Dezhang Li [1, b)], Xinzijian Liu [1, b)]

1. *Beijing National Laboratory for Molecular Sciences, Institute of Theoretical and*

   *Computational Chemistry, College of Chemistry and Molecular Engineering,*

   *Peking University, Beijing 100871, China*

a) Electronic mail: jianliupku@pku.edu.cn

b) Both authors contributed equally to the work

Below we extend the four integrators (BAOAB, BAOAB-num, OBABO, and OBABO-num) to normal mode path integral molecular dynamics and make the error analysis for the harmonic system and/or the free particle system.

**B1. The four integrators for normal mode path integral molecular dynamics**

The partition function of the canonical ensemble can be expressed in the coordinate space as Eq. (4). Consider the normal mode transformation of path integral beads[1-3]

$$\mathbf{x} = \mathbf{C}^{norm}\mathbf{q} \qquad (B1)$$

with $\mathbf{q} = \begin{pmatrix} \mathbf{q}_0 & \cdots & \mathbf{q}_{P-1} \end{pmatrix}^T$ and $\mathbf{x} = \begin{pmatrix} \mathbf{x}_1 & \cdots & \mathbf{x}_P \end{pmatrix}^T$, and the elements of the orthogonal transformation matrix $\mathbf{C}^{norm}$ are

$$C_{jk}^{norm} = \begin{cases} \sqrt{1/P}, & k = 0 \\ \sqrt{2/P}\cos\left(2\pi jk/P\right), & 1 \le k \le P/2 - 1 \\ \sqrt{1/P}\left(-1\right)^j, & k = P/2 \\ \sqrt{2/P}\sin\left(2\pi jk/P\right), & P/2 + 1 \le k \le P-1 \end{cases}, \qquad (B2)$$

for even $P$ and

$$C_{jk}^{norm} = \begin{cases} \sqrt{1/P}, & k = 0 \\ \sqrt{2/P}\cos\left(2\pi jk/P\right), & 1 \le k \le (P-1)/2 \\ \sqrt{2/P}\sin\left(2\pi jk/P\right), & (P+1)/2 \le k \le P-1 \end{cases}, \qquad (B3)$$

for odd $P$, with $\left(j = \overline{1, P}\right)$ in Eqs. (B2) and (B3). When the normal mode transformation of the path integral beads is employed, Eq. (4) becomes

$$Z = \lim_{P \to \infty} \left(\frac{P}{2\pi\beta\hbar^2}\right)^{NP/2} |\mathbf{M}|^{P/2} \int d\mathbf{q}_0 \int d\mathbf{q}_1 \cdots \int d\mathbf{q}_{P-1}$$
$$\times \exp\left\{-\beta\left[\sum_{k=0}^{P-1} \frac{1}{2}\omega_k^2 \mathbf{q}_k^T \overline{\mathbf{M}}_k^{norm}\mathbf{q}_k + \frac{1}{P}\sum_{j=1}^{P} V\left(\mathbf{x}_j\left(\mathbf{q}_0, \cdots, \mathbf{q}_{P-1}\right)\right)\right]\right\} \qquad .(B4)$$

If the (diagonal) mass matrices are chosen as $\bar{\mathbf{M}}_k^{norm} = \mathbf{M}$ $\left(k = \overline{0, P-1}\right)$, the frequency for each mode is given by

$$\omega_k = 2\omega_P \sin\left(k\pi / P\right) \quad \left(k = \overline{0, P-1}\right) \quad , \tag{B5}$$

respectively.  Eq. (16) then becomes

$$\phi\left(\mathbf{q}_0, \cdots, \mathbf{q}_{P-1}\right) = \frac{1}{P} \sum_{j=1}^{P} V\left(\mathbf{x}_j\left(\mathbf{q}_0, \cdots, \mathbf{q}_{P-1}\right)\right) \quad , \tag{B6}$$

and the term $\partial\phi / \partial\mathbf{q}_k$ is obtained from

$$\frac{\partial\phi}{\partial\mathbf{q}} = \left(\frac{\partial\mathbf{x}}{\partial\mathbf{q}}\right)^T \frac{\partial\phi}{\partial\mathbf{x}} = \left(\mathbf{C}^{norm}\right)^T \frac{\partial\phi}{\partial\mathbf{x}} \quad . \tag{B7}$$

Inserting fictitious momenta $\left(\mathbf{p}_0, \cdots, \mathbf{p}_{P-1}\right)$ into Eq. (B4) leads to

$$Z = \lim_{P \to \infty} \left(\frac{P}{4\pi^2\hbar^2}\right)^{NP/2} |\mathbf{M}|^{P/2} \left(\prod_{k=0}^{P-1} \left|\tilde{\mathbf{M}}_k^{norm}\right|\right)^{-1/2} \int\left(\prod_{k=0}^{P-1} d\mathbf{q}_k d\mathbf{p}_k\right) \quad ,$$
$$\times \exp\left[-\beta H_{eff}^{norm}\left(\mathbf{q}_0, \cdots, \mathbf{q}_{P-1}; \mathbf{p}_0, \cdots, \mathbf{p}_{P-1}\right)\right] \tag{B8}$$

with the effective Hamiltonian expressed by the normal mode variables

$$H_{eff}^{norm}(\mathbf{q}; \mathbf{p}) = \frac{1}{2}\mathbf{p}^T \tilde{\mathbf{M}}_{norm}^{-1} \mathbf{p} + U_{eff}^{norm}(\mathbf{q}) \tag{B9}$$

with

$$U_{eff}^{norm}(\mathbf{q}) = \frac{1}{2}\mathbf{q}^T \bar{\mathbf{M}}_{norm} \bar{\mathbf{\Omega}}^2 \mathbf{q} + \phi(\mathbf{q}) \quad . \tag{B10}$$

Here $\bar{\mathbf{M}}_{norm} = \begin{pmatrix} \mathbf{M}_{N\times N} & & \\ & \ddots & \\ & & \mathbf{M}_{N\times N} \end{pmatrix}$ and

$$\bar{\mathbf{\Omega}} = \begin{pmatrix} \omega_0 \mathbf{1}_{N\times N} & & \\ & \ddots & \\ & & \omega_{P-1}\mathbf{1}_{N\times N} \end{pmatrix} \quad . \tag{B11}$$

While the fictitious masses can be arbitrary, they are chosen as

$\tilde{\mathbf{M}}_{norm} = \begin{pmatrix} \mathbf{M}_{N\times N} & & \\ & \ddots & \\ & & \mathbf{M}_{N\times N} \end{pmatrix}$ for convenience in the present manuscript.

The thermodynamic property can then be evaluated using the following equation

$$\left\langle \hat{B} \right\rangle = \lim_{P \to \infty} \frac{\int \left( \prod_{k=0}^{P-1} d\mathbf{q}_k d\mathbf{p}_k \right) \exp\left\{ -\beta H_{eff}\left( \mathbf{q}_0, \cdots, \mathbf{q}_{P-1}; \mathbf{p}_0, \cdots, \mathbf{p}_{P-1} \right) \right\} \tilde{B}\left( \mathbf{x}_1, \cdots, \mathbf{x}_P \right)}{\int \left( \prod_{k=0}^{P-1} d\mathbf{q}_k d\mathbf{p}_k \right) \exp\left\{ -\beta H_{eff}\left( \mathbf{q}_0, \cdots, \mathbf{q}_{P-1}; \mathbf{p}_0, \cdots, \mathbf{p}_{P-1} \right) \right\}} \quad . \quad (B12)$$

A MD scheme coupled to a thermostatting method can be used to sample $\left( \mathbf{q}_0, \cdots, \mathbf{q}_{P-1}, \mathbf{p}_0, \cdots, \mathbf{p}_{P-1} \right)$ under a proper canonical distribution. This is noted normal mode PIMD. While the density distribution of normal mode PIMD is

$$\rho_{eq}(\mathbf{q}; \mathbf{p}) = \frac{1}{Z_{eff}^{norm}} \exp\left\{ -\beta H_{eff}^{norm}(\mathbf{q}; \mathbf{p}) \right\} \quad , \quad (B13)$$

the configurational distribution of normal mode PIMD is

$$\rho_{eq}^{config}(\mathbf{q}) = \frac{1}{Z_{norm}^{config}} \exp\left\{ -\beta U_{eff}^{norm}(\mathbf{q}) \right\} \quad , \quad (B14)$$

with $Z_{eff}^{norm}$ and $Z_{norm}^{config}$ the normalization constants. Note that only the latter is important in PIMD.

When a white noise Langevin dynamics is employed to thermostat the normal mode path integral variables $\left( \mathbf{q}_0, \cdots, \mathbf{q}_{P-1}, \mathbf{p}_0, \cdots, \mathbf{p}_{P-1} \right)$ in PIMD, one can obtain the equations of motion

$$\begin{pmatrix} \dot{\mathbf{q}}_k \\ \dot{\mathbf{p}}_k \end{pmatrix} = \begin{pmatrix} \mathbf{M}^{-1} \mathbf{p}_k \\ -\omega_k^2 \mathbf{M} \mathbf{q}_k - \dfrac{\partial \phi}{\partial \mathbf{q}_k} - \gamma_{norm}^{(k)} \mathbf{p}_k + \sqrt{\dfrac{2\gamma_{norm}^{(k)}}{\beta}} \mathbf{M}^{1/2} \mathbf{\eta}_k(t) \end{pmatrix} \quad \left( k = \overline{0, P-1} \right). \quad (B15)$$

Here $\mathbf{\eta}_k(t)$ and $\gamma_{norm}^{(k)}$ are similar to those in Eq. (24). According to Appendix I, the optimum value of the friction coefficient for the $k$-th normal mode is then

$$\gamma_{norm}^{(k)} = \omega_k \qquad \left( k = \overline{1, P-1} \right) \quad . \quad (B16)$$

Similar to Eq. (26), Eq. (B15) can be divided into three parts

$$\begin{pmatrix} \dot{\mathbf{q}}_k \\ \dot{\mathbf{p}}_k \end{pmatrix} = \underbrace{\begin{pmatrix} \mathbf{M}^{-1}\,\mathbf{p}_k \\ -\omega_k^2 \mathbf{M}\mathbf{q}_k \end{pmatrix}}_{A} + \underbrace{\begin{pmatrix} 0 \\ -\dfrac{\partial \phi}{\partial \mathbf{q}_k} \end{pmatrix}}_{B} + \underbrace{\begin{pmatrix} 0 \\ -\gamma_{norm}^{(k)}\mathbf{p}_k + \sqrt{\dfrac{2\gamma_{norm}^{(k)}}{\beta}}\,\mathbf{M}^{1/2}\boldsymbol{\eta}_k\left(t\right) \end{pmatrix}}_{O} \quad \left(k = \overline{0, P-1}\right). \text{(B17)}$$

The BAOAB integrator for propagating normal mode PIMD trajectory through a time interval $\Delta t$ for Eq. (B15) is

$$\mathbf{p}_k \leftarrow \mathbf{p}_k - \frac{\partial \phi}{\partial \mathbf{q}_k}\frac{\Delta t}{2} \qquad \left(k = \overline{0, P-1}\right) \tag{B18}$$

$$\mathbf{q}_0 \leftarrow \mathbf{q}_0 + \mathbf{M}^{-1}\mathbf{p}_0 \frac{\Delta t}{2}$$
$$\begin{pmatrix} \mathbf{q}_k \\ \mathbf{p}_k \end{pmatrix} \leftarrow \begin{pmatrix} \cos\left(\omega_k \Delta t/2\right)\mathbf{1} & \sin\left(\omega_k \Delta t/2\right)/\omega_k\,\mathbf{M}^{-1} \\ -\omega_k \sin\left(\omega_k \Delta t/2\right)\mathbf{M} & \cos\left(\omega_k \Delta t/2\right)\mathbf{1} \end{pmatrix}\begin{pmatrix} \mathbf{q}_k \\ \mathbf{p}_k \end{pmatrix} \quad \left(k = \overline{1, P-1}\right) \tag{B19}$$

$$\mathbf{p}_k \leftarrow \tilde{c}_{norm,1}^{(k)}\mathbf{p}_k + \tilde{c}_{norm,2}^{(k)}\sqrt{\frac{1}{\beta}}\mathbf{M}^{1/2}\boldsymbol{\eta}_k \qquad \left(k = \overline{0, P-1}\right) \tag{B20}$$

$$\mathbf{q}_0 \leftarrow \mathbf{q}_0 + \mathbf{M}^{-1}\mathbf{p}_0 \frac{\Delta t}{2}$$
$$\begin{pmatrix} \mathbf{q}_k \\ \mathbf{p}_k \end{pmatrix} \leftarrow \begin{pmatrix} \cos\left(\omega_k \Delta t/2\right)\mathbf{1} & \sin\left(\omega_k \Delta t/2\right)/\omega_k\,\mathbf{M}^{-1} \\ -\omega_k \sin\left(\omega_k \Delta t/2\right)\mathbf{M} & \cos\left(\omega_k \Delta t/2\right)\mathbf{1} \end{pmatrix}\begin{pmatrix} \mathbf{q}_k \\ \mathbf{p}_k \end{pmatrix} \quad \left(k = \overline{1, P-1}\right) \tag{B21}$$

$$\mathbf{p}_k \leftarrow \mathbf{p}_k - \frac{\partial \phi}{\partial \mathbf{q}_k}\frac{\Delta t}{2} \qquad \left(k = \overline{0, P-1}\right) \quad . \tag{B22}$$

Here the coefficients $\tilde{c}_{norm,1}^{(k)}$ and $\tilde{c}_{norm,2}^{(k)}$ are

$$\begin{aligned} \tilde{c}_{norm,1}^{(k)} &= \exp\left[-\gamma_{norm}^{(k)}\Delta t\right] \\ \tilde{c}_{norm,2}^{(k)} &= \sqrt{1 - \left(\tilde{c}_{norm,1}^{(k)}\right)^2} \end{aligned} \quad \left(k = \overline{0, P-1}\right) \tag{B23}$$

.

The OBABO algorithm for normal mode PIMD is the same as the PILE thermostat developed by Ceriotti *et al.*[4]. Such an OBABO or PILD algorithm for propagating the PIMD trajectory through a time interval $\Delta t$ for Eq. (B15) is

$$\mathbf{p}_k \leftarrow c_{norm,1}^{(k)}\mathbf{p}_k + c_{norm,2}^{(k)}\sqrt{\frac{1}{\beta}}\mathbf{M}^{1/2}\boldsymbol{\eta}_k \qquad \left(k=\overline{0,P-1}\right) \qquad (B24)$$

$$\mathbf{p}_k \leftarrow \mathbf{p}_k - \frac{\partial\phi}{\partial\mathbf{q}_k}\frac{\Delta t}{2} \qquad \left(k=\overline{0,P-1}\right) \qquad (B25)$$

$$\mathbf{q}_0 \leftarrow \mathbf{q}_0 + \mathbf{M}^{-1}\mathbf{p}_0\Delta t$$
$$\begin{pmatrix}\mathbf{q}_k\\\mathbf{p}_k\end{pmatrix} \leftarrow \begin{pmatrix}\cos\left(\omega_k\Delta t\right)\mathbf{1} & \sin\left(\omega_k\Delta t\right)/\omega_k\,\mathbf{M}^{-1}\\-\omega_k\sin\left(\omega_k\Delta t\right)\mathbf{M} & \cos\left(\omega_k\Delta t\right)\mathbf{1}\end{pmatrix}\begin{pmatrix}\mathbf{q}_k\\\mathbf{p}_k\end{pmatrix} \quad \left(k=\overline{1,P-1}\right) \quad (B26)$$

$$\mathbf{p}_k \leftarrow \mathbf{p}_k - \frac{\partial\phi}{\partial\mathbf{q}_k}\frac{\Delta t}{2} \qquad \left(k=\overline{0,P-1}\right) \qquad (B27)$$

$$\mathbf{p}_k \leftarrow c_{norm,1}^{(k)}\mathbf{p}_k + c_{norm,2}^{(k)}\sqrt{\frac{1}{\beta}}\mathbf{M}^{1/2}\boldsymbol{\eta}_k \qquad \left(k=\overline{0,P-1}\right)\ , \qquad (B28)$$

where the coefficients $c_{norm,1}^{(k)}$ and $c_{norm,2}^{(k)}$ are

$$c_{norm,1}^{(k)} = \exp\left[-\gamma_{norm}^{(k)}\Delta t/2\right]$$
$$c_{norm,2}^{(k)} = \sqrt{1-\left(c_{norm,1}^{(k)}\right)^2} \qquad \left(k=\overline{0,P-1}\right) \qquad , \qquad (B29)$$

respectively.

Langevin dynamics Eq. (B15) can also be divided into three parts in an alternative way

$$\begin{pmatrix}\dot{\mathbf{q}}_k\\\dot{\mathbf{p}}_k\end{pmatrix} = \underbrace{\begin{pmatrix}\mathbf{M}^{-1}\mathbf{p}_k\\0\end{pmatrix}}_{A} + \underbrace{\begin{pmatrix}0\\-\omega_k^2\mathbf{M}\mathbf{q}_k - \dfrac{\partial\phi}{\partial\mathbf{q}_k}\end{pmatrix}}_{B}$$
$$+ \underbrace{\begin{pmatrix}0\\-\gamma_{norm}^{(k)}\mathbf{p}_k + \sqrt{\dfrac{2\gamma_{norm}^{(k)}}{\beta}}\,\mathbf{M}^{1/2}\boldsymbol{\eta}_k\left(t\right)\end{pmatrix}}_{O} \qquad \left(k=\overline{0,P-1}\right)$$
$$. \qquad (B30)$$

That is, the force term $-\omega_k^2\mathbf{M}\mathbf{q}_k$ is moved from part A to part B in Eq. (B17). Here the analytical propagator of the harmonic part of Eq. (B17) is no longer used. The BAOAB-num integrator for propagating normal mode PIMD trajectory through a time interval $\Delta t$ for Eq. (B30) is

$$\mathbf{p}_k \leftarrow \mathbf{p}_k - \frac{\partial \phi}{\partial \mathbf{q}_k} \frac{\Delta t}{2} - \omega_k^2 \mathbf{M} \mathbf{q}_k \frac{\Delta t}{2} \qquad \left( k = \overline{0, P-1} \right) \tag{B31}$$

$$\mathbf{q}_k \leftarrow \mathbf{q}_k + \mathbf{M}^{-1} \mathbf{p}_k \frac{\Delta t}{2} \qquad \left( k = \overline{0, P-1} \right) \tag{B32}$$

$$\mathbf{p}_k \leftarrow \tilde{c}_{norm,1}^{(k)} \mathbf{p}_k + \tilde{c}_{norm,2}^{(k)} \sqrt{\frac{1}{\beta}} \mathbf{M}^{1/2} \mathbf{\eta}_k \qquad \left( k = \overline{0, P-1} \right) \tag{B33}$$

$$\mathbf{q}_k \leftarrow \mathbf{q}_k + \mathbf{M}^{-1} \mathbf{p}_k \frac{\Delta t}{2} \qquad \left( k = \overline{0, P-1} \right) \tag{B34}$$

$$\mathbf{p}_k \leftarrow \mathbf{p}_k - \frac{\partial \phi}{\partial \mathbf{q}_k} \frac{\Delta t}{2} - \omega_k^2 \mathbf{M} \mathbf{q}_k \frac{\Delta t}{2} \qquad \left( k = \overline{0, P-1} \right) \ . \tag{B35}$$

Here $\mathbf{\eta}_k$, $\tilde{c}_{norm,1}^{(k)}$ and $\tilde{c}_{norm,2}^{(k)}$ are defined in the same way as in BAOAB [Eq. (B18) -(B22)].

Similarly, the OBABO-num integrator for propagating normal mode PIMD trajectory through a time interval $\Delta t$ for Eq. (B30) is

$$\mathbf{p}_k \leftarrow c_{norm,1}^{(k)} \mathbf{p}_k + c_{norm,2}^{(k)} \sqrt{\frac{1}{\beta}} \mathbf{M}^{1/2} \mathbf{\eta}_k \qquad \left( k = \overline{0, P-1} \right) \tag{B36}$$

$$\mathbf{p}_k \leftarrow \mathbf{p}_k - \frac{\partial \phi}{\partial \mathbf{q}_k} \frac{\Delta t}{2} - \omega_k^2 \mathbf{M} \mathbf{q}_k \frac{\Delta t}{2} \qquad \left( k = \overline{0, P-1} \right) \tag{B37}$$

$$\mathbf{q}_k \leftarrow \mathbf{q}_k + \mathbf{M}^{-1} \mathbf{p}_k \Delta t \qquad \left( k = \overline{0, P-1} \right) \tag{B38}$$

$$\mathbf{p}_k \leftarrow \mathbf{p}_k - \frac{\partial \phi}{\partial \mathbf{q}_k} \frac{\Delta t}{2} - \omega_k^2 \mathbf{M} \mathbf{q}_k \frac{\Delta t}{2} \qquad \left( k = \overline{0, P-1} \right) \tag{B39}$$

$$\mathbf{p}_k \leftarrow c_{norm,1}^{(k)} \mathbf{p}_k + c_{norm,2}^{(k)} \sqrt{\frac{1}{\beta}} \mathbf{M}^{1/2} \mathbf{\eta}_k \qquad \left( k = \overline{0, P-1} \right) \ . \tag{B40}$$

Here $\mathbf{\eta}_k$, $c_{norm,1}^{(k)}$ and $c_{norm,2}^{(k)}$ are defined in the same way as in OBABO [Eq. (B24) -(B28)].

## B2. Accuracy of the normal mode PIMD integrators

Below we investigate the accuracy of BAOAB, OBABO, BAOAB-num and OBABO-num for normal mode PIMD for the harmonic system and for the free particle system.

## 1. The harmonic system

Consider the harmonic system [Eq. (70)]. Define the symmetric positive-definite matrix $\mathbf{K}^{norm} = \dfrac{1}{P}\mathbf{A}$, then Eq. (B6) becomes

$$
\begin{aligned}
\phi(\mathbf{q}) &= \frac{1}{2}\left(\mathbf{x} - \overline{\mathbf{x}}_{eq}\right)^T \overline{\mathbf{K}}\left(\mathbf{x} - \overline{\mathbf{x}}_{eq}\right) \\
&= \frac{1}{2}\left(\mathbf{q} - \mathbf{q}_{eq}\right)^T \left(\overline{\mathbf{C}}^{norm}\right)^T \overline{\mathbf{K}}\overline{\mathbf{C}}^{norm}\left(\mathbf{q} - \mathbf{q}_{eq}\right)
\end{aligned}
\tag{B41}
$$

where $\quad \overline{\mathbf{C}}^{norm} = \begin{pmatrix} C_{10}^{norm}\mathbf{1}_{N\times N} & \cdots & C_{1(P-1)}^{norm}\mathbf{1}_{N\times N} \\ \vdots & \ddots & \vdots \\ C_{P0}^{norm}\mathbf{1}_{N\times N} & \cdots & C_{P(P-1)}^{norm}\mathbf{1}_{N\times N} \end{pmatrix} \quad, \qquad \overline{\mathbf{K}} = \begin{pmatrix} \mathbf{K}_{N\times N}^{norm} & & \\ & \ddots & \\ & & \mathbf{K}_{N\times N}^{norm} \end{pmatrix}$ ,

$\overline{\mathbf{x}}_{eq} = \left(\mathbf{x}_{eq} \quad \cdots \quad \mathbf{x}_{eq}\right)^T$, and

$$
\mathbf{q}_{eq} = \left(\sqrt{P}\mathbf{x}_{eq} \quad 0 \quad \cdots \quad 0\right)^T \qquad .
\tag{B42}
$$

It is trivial to prove

$$
\phi(\mathbf{q}) = \frac{1}{2}\left(\mathbf{q} - \mathbf{q}_{eq}\right)^T \overline{\mathbf{K}}\left(\mathbf{q} - \mathbf{q}_{eq}\right)
\tag{B43}
$$

by virtue of the characteristic of the orthogonal matrix. Substituting Eq. (B43) into Eq. (B10), one obtains

$$
U_{eff}^{norm}(\mathbf{q}) = \frac{1}{2}\mathbf{q}^T \overline{\mathbf{M}}_{norm}\overline{\mathbf{\Omega}}^2\mathbf{q} + \frac{1}{2}\left(\mathbf{q} - \mathbf{q}_{eq}\right)^T \overline{\mathbf{K}}\left(\mathbf{q} - \mathbf{q}_{eq}\right) \quad .
\tag{B44}
$$

Substitute Eqs. (B5), (B11), and (B42) into Eq. (B44), one finds

$$
\begin{aligned}
U_{eff}^{norm}(\mathbf{q}) = &\frac{1}{2}\left(\mathbf{q}_0 - \sqrt{P}\mathbf{x}_{eq}\right)^T \mathbf{K}^{norm}\left(\mathbf{q}_0 - \sqrt{P}\mathbf{x}_{eq}\right) \\
&+ \sum_{k=1}^{P-1}\left[\frac{1}{2}\omega_k^2\mathbf{q}_k^T\mathbf{M}\mathbf{q}_k + \frac{1}{2}\mathbf{q}_k^T\mathbf{K}^{norm}\mathbf{q}_k\right]
\end{aligned}
\tag{B45}
$$

The effective Hamiltonian Eq. (B9) then becomes

$$H_{eff}^{norm}\left(\mathbf{q};\mathbf{p}\right) = \sum_{k=0}^{P-1} H_k^{norm}\left(\mathbf{q}_k;\mathbf{p}_k\right) \tag{B46}$$

with

$$H_0^{norm}\left(\mathbf{q}_0;\mathbf{p}_0\right) = \frac{1}{2}\mathbf{p}_0^T\mathbf{M}^{-1}\mathbf{p}_0 + \frac{1}{2}\left(\mathbf{q}_0 - \sqrt{P}\mathbf{x}_{eq}\right)^T\mathbf{K}^{norm}\left(\mathbf{q}_0 - \sqrt{P}\mathbf{x}_{eq}\right) \tag{B47}$$

and

$$H_k^{norm}\left(\mathbf{q}_k;\mathbf{p}_k\right) = \frac{1}{2}\mathbf{p}_k^T\mathbf{M}^{-1}\mathbf{p}_k + \frac{1}{2}\omega_k^2\mathbf{q}_k^T\mathbf{M}\mathbf{q}_k + \frac{1}{2}\mathbf{q}_k^T\mathbf{K}^{norm}\mathbf{q}_k \quad\left(k = \overline{1,P-1}\right) \ . \tag{B48}$$

The normal modes are independent of one another. Eq. (B13) and Eq. (B14) then become

$$\rho_{eq}\left(\mathbf{q};\mathbf{p}\right) = \prod_{k=0}^{P-1}\frac{1}{Z_k^{norm}}\exp\left\{-\beta H_k^{norm}\left(\mathbf{q}_k;\mathbf{p}_k\right)\right\} \tag{B49}$$

and

$$\rho_{eq}^{config}\left(\mathbf{q}\right) = \frac{1}{Z_0^{config}}\exp\left\{-\beta\left[\frac{1}{2}\left(\mathbf{q}_0 - \sqrt{P}\mathbf{x}_{eq}\right)^T\mathbf{K}^{norm}\left(\mathbf{q}_0 - \sqrt{P}\mathbf{x}_{eq}\right)\right]\right\}$$
$$\times\prod_{k=1}^{P-1}\frac{1}{Z_k^{config}}\exp\left\{-\beta\left[\frac{1}{2}\omega_k^2\mathbf{q}_k^T\mathbf{M}\mathbf{q}_k + \frac{1}{2}\mathbf{q}_k^T\mathbf{K}^{norm}\mathbf{q}_k\right]\right\} \ , \tag{B50}$$

respectively. Here $\left\{Z_k^{norm}\right\}$ and $\left\{Z_k^{config}\right\}\left(k = \overline{0,P-1}\right)$ are the normalization coefficients. Below we discuss the accuracy of the four algorithms, respectively.

(1) BAOAB

The propagation of each mode is independent of one another. By virtue of the conclusion in Appendix Ⅲ and Eq. (B47), the steady density of the 0-th normal mode (or the centroid mode) is

$$\rho_0^{BAOAB}\left(\mathbf{q}_0;\mathbf{p}_0\right) = \frac{1}{Z_0'}\exp\left\{-\beta\left[\frac{1}{2}\mathbf{p}_0^T\left(\mathbf{M} - \mathbf{K}^{norm}\frac{\Delta t^2}{4}\right)^{-1}\mathbf{p}_0\right.\right.$$
$$\left.\left. + \frac{1}{2}\left(\mathbf{q}_0 - \sqrt{P}\mathbf{x}_{eq}\right)^T\mathbf{K}^{norm}\left(\mathbf{q}_0 - \sqrt{P}\mathbf{x}_{eq}\right)\right]\right\} \ , \tag{B51}$$

where $Z_0'$ is the new normalization coefficient. Integration over $\mathbf{p}_0$ in Eq. (B51) leads to the exact configurational distribution for the $0^{th}$ normal mode. Langevin dynamics for the $k$-th mode $\left(k = \overline{1, P-1}\right)$ [Eq. (B17)] becomes

$$\begin{pmatrix} \dot{\mathbf{q}}_k \\ \dot{\mathbf{p}}_k \end{pmatrix} = \underbrace{\begin{pmatrix} \mathbf{M}^{-1}\mathbf{p}_k \\ -\omega_k^2 \mathbf{M}\mathbf{q}_k \end{pmatrix}}_{A} + \underbrace{\begin{pmatrix} 0 \\ -\mathbf{K}^{norm}\mathbf{q}_k \end{pmatrix}}_{B} + \underbrace{\begin{pmatrix} 0 \\ -\gamma_{norm}^{(k)}\mathbf{p}_k + \sqrt{\dfrac{2\gamma_{norm}^{(k)}}{\beta}}\,\mathbf{M}^{1/2}\mathbf{\eta}_k\left(t\right) \end{pmatrix}}_{O} . \quad \text{(B52)}$$

The phase space propagator for the $k$-th mode $\left(k = \overline{1, P-1}\right)$ of BAOAB during the time interval $\Delta t$ is

$$e^{\mathcal{L}_{BAOAB}^{(k)}\Delta t} = e^{\mathcal{L}_B^{(k)}\Delta t/2}e^{\mathcal{L}_A^{(k)}\Delta t/2}e^{\mathcal{L}_O^{(k)}\Delta t}e^{\mathcal{L}_A^{(k)}\Delta t/2}e^{\mathcal{L}_B^{(k)}\Delta t/2} \quad , \quad \text{(B53)}$$

where

$$\mathcal{L}_A^{(k)}\rho = -\mathbf{p}_k^T\mathbf{M}^{-1}\frac{\partial\rho}{\partial\mathbf{q}_k} + \omega_k^2\mathbf{q}_k^T\mathbf{M}\frac{\partial\rho}{\partial\mathbf{p}_k} \quad , \quad \text{(B54)}$$

$$\mathcal{L}_B^{(k)}\rho = \mathbf{q}_k^T\mathbf{K}^{norm}\frac{\partial\rho}{\partial\mathbf{p}_k} \quad , \quad \text{(B55)}$$

$$\mathcal{L}_O^{(k)}\rho = -\frac{\partial}{\partial\mathbf{p}_k}\cdot\left(\gamma_{norm}^{(k)}\mathbf{p}_k\rho\right) + \frac{1}{\beta}\frac{\partial}{\partial\mathbf{p}_k}\cdot\left(\gamma_{norm}^{(k)}\mathbf{M}\frac{\partial\rho}{\partial\mathbf{p}_k}\right) \quad . \quad \text{(B56)}$$

It is straightforward to show that

$$e^{\mathcal{L}_A^{(k)}\Delta t/2}f\left(\mathbf{q}_k,\mathbf{p}_k\right) = f\left(\cos\left(\omega_k\Delta t/2\right)\mathbf{q}_k - \sin\left(\omega_k\Delta t/2\right)/\omega_k\mathbf{M}^{-1}\mathbf{p}_k, \right. \\ \left. \cos\left(\omega_k\Delta t/2\right)\mathbf{p}_k + \omega_k\sin\left(\omega_k\Delta t/2\right)\mathbf{M}\mathbf{q}_k\right) \quad , \quad \text{(B57)}$$

$$e^{\mathcal{L}_B^{(k)}\Delta t/2}g\left(\mathbf{q}_k,\mathbf{p}_k\right) = g\left(\mathbf{q}_k,\mathbf{p}_k + \mathbf{K}^{norm}\mathbf{q}_k\frac{\Delta t}{2}\right) \quad , \quad \text{(B58)}$$

and the OU process keeps the Maxwell momentum distribution unchanged, i.e.,

$$e^{\mathcal{L}_O^{(k)}\Delta t}\exp\left\{-\beta\left[\frac{1}{2}\mathbf{p}_k^T\mathbf{M}^{-1}\mathbf{p}_k\right]\right\} = \exp\left\{-\beta\left[\frac{1}{2}\mathbf{p}_k^T\mathbf{M}^{-1}\mathbf{p}_k\right]\right\} \quad . \quad \text{(B59)}$$

Consider the density distribution

$$\rho_k^{BAOAB}\left(\mathbf{q}_k;\mathbf{p}_k\right)=\frac{1}{Z_k'}\exp\left\{-\frac{\beta}{2}\mathbf{q}_k^T\left[\omega_k^2\mathbf{M}+\omega_k\frac{\Delta t}{2}\cot\left(\omega_k\frac{\Delta t}{2}\right)\mathbf{K}^{norm}\right]\mathbf{q}_k\right.$$
$$\left.-\frac{\beta}{2}\mathbf{p}_k^T\left[\mathbf{M}-\frac{1}{\omega_k}\frac{\Delta t}{2}\tan\left(\omega_k\frac{\Delta t}{2}\right)\mathbf{K}^{norm}\right]^{-1}\mathbf{p}_k\right\}\quad, \quad(B60)$$

where $Z_k'$ is the new normalization coefficient. Using Eqs. (B53) and (B57)-(B59), it is easy to verify

$$e^{\mathcal{L}_{BAOAB}^{(k)}\Delta t}\rho_k^{BAOAB}=\rho_k^{BAOAB}\quad. \quad(B61)$$

That is, Eq. (B60) is a steady state of the BAOAB integrator. Multiplying Eq. (B51) and Eq. (B60), i.e.,

$$\rho_{eq,norm}^{BAOAB}\left(\mathbf{q};\mathbf{p}\right)=\prod_{k=0}^{P-1}\rho_k^{BAOAB}\left(\mathbf{q}_k;\mathbf{p}_k\right)\quad, \quad(B62)$$

and then integrating over $\mathbf{p}_k$ $\left(k=\overline{0,P-1}\right)$ produce the configurational distribution for BAOAB

$$\rho_{BAOAB}^{config,norm}\left(\mathbf{q}\right)=\frac{1}{Z_0^{config}}\exp\left\{-\frac{\beta}{2}\left(\mathbf{q}_0-\sqrt{P}\mathbf{x}_{eq}\right)^T\mathbf{K}^{norm}\left(\mathbf{q}_0-\sqrt{P}\mathbf{x}_{eq}\right)\right\}$$
$$\times\prod_{k=1}^{P-1}\frac{1}{Z_k''}\exp\left\{-\frac{\beta}{2}\mathbf{q}_k^T\left[\omega_k^2\mathbf{M}+\omega_k\frac{\Delta t}{2}\cot\left(\omega_k\frac{\Delta t}{2}\right)\mathbf{K}^{norm}\right]\mathbf{q}_k\right\}\quad. \quad(B63)$$

Expanding Eq. (B63) into a power series of $\Delta t$, one obtains

$$\rho_{BAOAB}^{config,norm}\left(\mathbf{q}\right)=\rho_{eq}^{config}\left(\mathbf{q}\right)\left[1+\frac{\beta\Delta t^2}{24}\sum_{k=1}^{P-1}\omega_k^2\mathbf{q}_k^T\mathbf{K}^{norm}\mathbf{q}_k\right.$$
$$\left.-\frac{\Delta t^2}{24}\sum_{k=1}^{P-1}\omega_k^2\text{Tr}\left[\left(\omega_k^2\mathbf{M}+\mathbf{K}^{norm}\right)^{-1}\mathbf{K}^{norm}\right]+O\left(\Delta t^4\right)\right]. \quad(B64)$$

Compare Eq. (B64) of normal mode PIMD to Eq. (79) of staging PIMD. One finds that the error of the configurational distribution of BAOAB for normal mode PIMD is close to that for staging PIMD in the harmonic limit, when one considers the substitutions $\omega_P\to\omega_k$ and $\left(\mathbf{S}^{-1}\right)^T\mathbf{S}^{-1}\to\mathbf{1}$.

(2) OBABO

Similarly, the steady density for the OBABO integrator is

$$\rho_{eq,\,norm}^{OBABO}\left(\mathbf{q};\mathbf{p}\right)=\frac{1}{\overline{Z}_0}\exp\left\{-\frac{\beta}{2}\mathbf{p}_0^T\mathbf{M}^{-1}\mathbf{p}_0\right.$$
$$\left.-\frac{\beta}{2}\left(\mathbf{q}_0-\sqrt{P}\mathbf{x}_{eq}\right)^T\left(\mathbf{1}-\mathbf{K}^{norm}\mathbf{M}^{-1}\frac{\Delta t^2}{4}\right)\mathbf{K}^{norm}\left(\mathbf{q}_0-\sqrt{P}\mathbf{x}_{eq}\right)\right]\right\}$$
$$\times\prod_{k=1}^{P-1}\frac{1}{\overline{\overline{Z}}_k}\exp\left\{-\frac{\beta}{2}\mathbf{p}_k^T\mathbf{M}^{-1}\mathbf{p}_k\right.$$
$$\left.-\frac{\beta}{2}\mathbf{q}_k^T\left[\omega_k^2\mathbf{M}+\omega_k\Delta t\cot\left(\omega_k\Delta t\right)\mathbf{K}^{norm}-\mathbf{K}^{norm}\mathbf{M}^{-1}\mathbf{K}^{norm}\frac{\Delta t^2}{4}\right]\mathbf{q}_k\right\}$$

.(B65)

Eq. (B65) produces the configurational distribution

$$\rho_{OBABO}^{config,\,norm}(\mathbf{q})=\frac{1}{\overline{\overline{Z}}_0}\exp\left\{-\frac{\beta}{2}\left(\mathbf{q}_0-\sqrt{P}\mathbf{x}_{eq}\right)^T\left(\mathbf{1}-\mathbf{K}^{norm}\mathbf{M}^{-1}\frac{\Delta t^2}{4}\right)\mathbf{K}^{norm}\left(\mathbf{q}_0-\sqrt{P}\mathbf{x}_{eq}\right)\right\}$$
$$\times\prod_{k=1}^{P-1}\frac{1}{\overline{\overline{Z}}_k}\exp\left\{-\frac{\beta}{2}\mathbf{q}_k^T\left[\omega_k^2\mathbf{M}+\omega_k\Delta t\cot\left(\omega_k\Delta t\right)\mathbf{K}^{norm}-\mathbf{K}^{norm}\mathbf{M}^{-1}\mathbf{K}^{norm}\frac{\Delta t^2}{4}\right]\mathbf{q}_k\right\}$$

.(B66)

Expanding the density into a power series of $\Delta t$, one obtains

$$\rho_{OBABO}^{config,\,norm}\left(\mathbf{q}\right)=\rho_{eq}^{config}\left(\mathbf{q}\right)\left[1+\frac{\beta\Delta t^2}{6}\sum_{k=1}^{P-1}\omega_k^2\mathbf{q}_k^T\mathbf{K}^{norm}\mathbf{q}_k+\frac{\beta\Delta t^2}{8}\sum_{k=1}^{P-1}\mathbf{q}_k^T\mathbf{K}^{norm}\mathbf{M}^{-1}\mathbf{K}^{norm}\mathbf{q}_k\right.$$
$$+\frac{\beta\Delta t^2}{8}\left(\mathbf{q}_0-\sqrt{P}\mathbf{x}_{eq}\right)^T\mathbf{K}^{norm}\mathbf{M}^{-1}\mathbf{K}^{norm}\left(\mathbf{q}_0-\sqrt{P}\mathbf{x}_{eq}\right)-\frac{P\Delta t^2}{8}\text{Tr}\left[\mathbf{M}^{-1}\mathbf{K}^{norm}\right]$$
$$\left.-\frac{\Delta t^2}{24}\sum_{k=1}^{P-1}\omega_k^2\text{Tr}\left[\left(\omega_k^2\mathbf{M}+\mathbf{K}^{norm}\right)^{-1}\mathbf{K}^{norm}\right]+O\left(\Delta t^4\right)\right]$$

(B67)

The error of the configurational distribution produced by OBABO is governed by the term $\dfrac{\beta\Delta t^2}{6}\sum_{k=1}^{P-1}\omega_k^2\mathbf{q}_k^T\mathbf{K}^{norm}\mathbf{q}_k$ for converged PIMD results. This error is close to that

of staging PIMD [Eq. (81)] when one considers the substitutions $\omega_P\rightarrow\omega_k$ and

$\left(\mathbf{S}^{-1}\right)^T\mathbf{S}^{-1}\rightarrow\mathbf{1}$. In comparison to BAOAB [Eq. (B64)], the error in OBABO [Eq.

(B67)] is increased by approximately a factor of 4 for the harmonic system when normal mode PIMD is employed.  The conclusion is the same as that for staging PIMD in Section II-4.

(3) BAOAB-num

Eq.  (B45)  can be expressed in a more compact form

$$U_{eff}^{norm}\left(\mathbf{q}\right)=\sum_{k=0}^{P-1}U_k^{eff}\left(\mathbf{q}_k\right) \tag{B68}$$

with

$$U_0^{eff}\left(\mathbf{q}_0\right)=\frac{1}{2}\left(\mathbf{q}_0-\sqrt{P}\mathbf{x}_{eq}\right)^T\mathbf{K}^{norm}\left(\mathbf{q}_0-\sqrt{P}\mathbf{x}_{eq}\right) \ , \tag{B69}$$

and

$$U_k^{eff}\left(\mathbf{q}_k\right)=\frac{1}{2}\mathbf{q}_k^T\left(\omega_k^2\mathbf{M}+\mathbf{K}^{norm}\right)\mathbf{q}_k, \quad \left(k=\overline{1,P-1}\right) \ . \tag{B70}$$

Eq.  (B30)  for Langevin dynamics then becomes

$$\begin{pmatrix}\dot{\mathbf{q}}_k\\\dot{\mathbf{p}}_k\end{pmatrix}=\underbrace{\begin{pmatrix}\mathbf{M}^{-1}\mathbf{p}_k\\0\end{pmatrix}}_{A}+\underbrace{\begin{pmatrix}0\\-\dfrac{\partial U_k^{eff}\left(\mathbf{q}_k\right)}{\partial\mathbf{q}_k}\end{pmatrix}}_{B}$$
$$+\underbrace{\begin{pmatrix}0\\-\gamma_{norm}^{(k)}\mathbf{p}_k+\sqrt{\dfrac{2\gamma_{norm}^{(k)}}{\beta}}\mathbf{M}^{1/2}\mathbf{\eta}_k\left(t\right)\end{pmatrix}}_{O} \quad \left(k=\overline{0,P-1}\right) \tag{B71}$$

Following the procedure for obtaining the steady density of BAOAB-num for staging PIMD in Appendix III, it is straightforward to verify that the configurational distribution given by BAOAB-num is

$$\rho_{BAOAB-num}^{config,norm}\left(\mathbf{q}\right)=\rho_{eq}^{config}\left(\mathbf{q}\right) \ . \tag{B72}$$

That is, the BAOAB-num integrator for normal mode PIMD in principle leads to the exact configurational distribution (of beads) for the harmonic system.   It is consistent with the conclusion for staging PIMD in Appendix III.

(4) OBABO–num

Similarly, the steady density distribution for the OBABO-num integrator is

$$
\begin{aligned}
\rho_{eq,norm}^{OBABO-num}\left(\mathbf{q};\mathbf{p}\right) = \frac{1}{\tilde{Z}_0}\exp\Bigg\{ &-\frac{\beta}{2}\mathbf{p}_0^T\mathbf{M}^{-1}\mathbf{p}_0 \\
&-\frac{\beta}{2}\left(\mathbf{q}_0-\sqrt{P}\mathbf{x}_{eq}\right)^T\left(\mathbf{1}-\mathbf{K}^{norm}\mathbf{M}^{-1}\frac{\Delta t^2}{4}\right)\mathbf{K}^{norm}\left(\mathbf{q}_0-\sqrt{P}\mathbf{x}_{eq}\right)\Bigg]\Bigg\} \\
&\times\prod_{k=1}^{P-1}\frac{1}{\tilde{Z}_k}\exp\Bigg\{ -\frac{\beta}{2}\mathbf{p}_k^T\mathbf{M}^{-1}\mathbf{p}_k \\
&-\frac{\beta}{2}\mathbf{q}_k^T\left[\mathbf{1}-\left(\omega_k^2\mathbf{M}+\mathbf{K}^{norm}\right)\mathbf{M}^{-1}\frac{\Delta t^2}{4}\right]\left(\omega_k^2\mathbf{M}+\mathbf{K}^{norm}\right)\mathbf{q}_k\Bigg\}
\end{aligned}
\tag{B73}
$$

which produces the configurational distribution

$$
\begin{aligned}
\rho_{OBABO-num}^{config,norm}\left(\mathbf{q}\right) = \frac{1}{\tilde{Z}_0}\exp\Bigg\{ &-\frac{\beta}{2}\left(\mathbf{q}_0-\sqrt{P}\mathbf{x}_{eq}\right)^T\left(\mathbf{1}-\mathbf{K}^{norm}\mathbf{M}^{-1}\frac{\Delta t^2}{4}\right)\mathbf{K}^{norm}\left(\mathbf{q}_0-\sqrt{P}\mathbf{x}_{eq}\right)\Bigg\} \\
&\times\prod_{k=1}^{P-1}\frac{1}{\tilde{Z}_k}\exp\Bigg\{-\frac{\beta}{2}\mathbf{q}_k^T\left[\mathbf{1}-\left(\omega_k^2\mathbf{M}+\mathbf{K}^{norm}\right)\mathbf{M}^{-1}\frac{\Delta t^2}{4}\right]\left(\omega_k^2\mathbf{M}+\mathbf{K}^{norm}\right)\mathbf{q}_k\Bigg\}
\end{aligned}
\tag{B74}
$$

One expands the density into a power series of $\Delta t$ and then obtains

$$
\begin{aligned}
\rho_{OBABO-num}^{config,norm}\left(\mathbf{q}\right) = \rho_{eq}^{config}\left(\mathbf{q}\right)\Bigg[ &1+\frac{\beta\Delta t^2}{8}\sum_{k=1}^{P-1}\omega_k^4\mathbf{q}_k^T\mathbf{M}\mathbf{q}_k+\frac{\beta\Delta t^2}{4}\sum_{k=1}^{P-1}\omega_k^2\mathbf{q}_k^T\mathbf{K}^{norm}\mathbf{q}_k \\
&+\frac{\beta\Delta t^2}{8}\sum_{k=1}^{P-1}\mathbf{q}_k^T\mathbf{K}^{norm}\mathbf{M}^{-1}\mathbf{K}^{norm}\mathbf{q}_k \\
&+\frac{\beta\Delta t^2}{8}\left(\mathbf{q}_0-\sqrt{P}\mathbf{x}_{eq}\right)^T\mathbf{K}^{norm}\mathbf{M}^{-1}\mathbf{K}^{norm}\left(\mathbf{q}_0-\sqrt{P}\mathbf{x}_{eq}\right) \\
&-\frac{P\Delta t^2}{8}\mathrm{Tr}\left[\mathbf{M}^{-1}\mathbf{K}^{norm}\right]-\frac{N\Delta t^2}{8}\sum_{k=1}^{P-1}\omega_k^2+O\left(\Delta t^4\right)\Bigg]
\end{aligned}
\tag{B75}
$$

The error of the configurational distribution produced by OBABO-num is governed by the term $\dfrac{\beta\Delta t^2}{8}\sum_{k=1}^{P-1}\omega_k^4\mathbf{q}_k^T\mathbf{M}\mathbf{q}_k$ for converged PIMD results. This error is close to that of staging PIMD [Eq. (A39)] when one considers the substitutions $\omega_P\to\omega_k$ and $\left(\mathbf{S}^{-1}\right)^T \mathbf{S}^{-1} \to \mathbf{1}$.   Comparing Eq. (B75) to Eqs. (B64), (B67), and (B72) for normal mode PIMD, one finds Eq. (A40) is also the ascending order for the error of the configurational distribution (in the harmonic limit).   It is the same as the conclusion for staging PIMD in Appendices II and III.

### 2.   The free particle system

Consider a free particle system where $\phi = 0$ or equivalently

$$\mathbf{K}^{norm} = 0 \qquad\qquad (B76)$$

in Eq. (B45).   For BAOAB and OBABO, Eq. (B63) and Eq. (B66) lead to the exact configurational distribution of the path integral beads. I.e., both BAOAB and OBABO are exact in the free particle limit.   BAOAB-num produces the exact configurational distribution for the harmonic system that includes the free particle limit.   Substituting Eq. (B76) into Eq. (B74), one finds that OBABO-num does *not* lead to the exact configurational distribution of the path integral beads even in the free particle limit. The conclusion for normal mode PIMD agrees with that in Appendix III when staging PIMD is used.